1
  2
  3
  4
  5
  6
  7
  8
  9
 10
 11
 12
 13
 14
 15
 16
 17
 18
 19
 20
 21
 22
 23
 24
 25
 26
 27
 28
 29
 30
 31
 32
 33
 34
 35
 36
 37
 38
 39
 40
 41
 42
 43
 44
 45
 46
 47
 48
 49
 50
 51
 52
 53
 54
 55
 56
 57
 58
 59
 60
 61
 62
 63
 64
 65
 66
 67
 68
 69
 70
 71
 72
 73
 74
 75
 76
 77
 78
 79
 80
 81
 82
 83
 84
 85
 86
 87
 88
 89
 90
 91
 92
 93
 94
 95
 96
 97
 98
 99
100
101
102
103
104
105
106
107
108
109
110
111
112
113
114
115
116
117
118
119
120
121
122
123
124
125
126
127
128
129
130
131
132
133
134
135
136
137
138
139
140
141
142
143
144
145
146
147
148
149
150
151
152
153
154
155
156
157
158
159
160
161
162
163
164
165
166
167
168
169
170
171
172
173
174
175
176
177
178
179
180
181
182
183
184
185
186
187
188
189
190
191
192
193
194
195
196
197
198
199
200
201
202
203
204
205
206
207
208
209
210
211
212
213
214
215
216
217
218
219
220
221
222
223
224
225
226
227
228
229
230
231
232
233
234
235
236
237
238
239
240
241
242
243
244
245
246
247
248
249
250
251
252
253
254
255
256
257
258
259
260
261
262
263
264
265
266
267
268
269
270
271
272
273
274
275
276
277
278
279
280
281
282
283
284
285
286
287
288
289
290
291
292
293
294
295
296
297
298
299
300
301
302
303
304
305
306
307
308
309
310
311
312
313
314
315
316
317
318
319
320
321
322
323
324
325
326
327
328
329
330
331
332
333
334
335
336
337
338
339
340
341
342
343
344
345
346
347
348
349
350
351
352
353
354
355
356
357
358
359
360
361
362
363
364
365
366
367
368
369
370
371
372
373
374
375
376
377
378
379
380
381
382
383
384
385
386
387
388
389
390
391
392
393
394
395
396
\documentclass[%
amsmath,amssymb,aps,showkeys,
prl,twocolumn
]{revtex4-1}

\usepackage{graphicx}
\usepackage{amsmath,bm}
\usepackage{amsmath}
\usepackage{amssymb}
\usepackage{graphicx}\usepackage{epsfig}
\usepackage{setspace}
\usepackage{color}
\usepackage{float}

\newcommand{\be}{\begin{equation}}
\newcommand{\ee}{\end{equation}}

\newcommand{\bea}{\begin{eqnarray}}
\newcommand{\eea}{\end{eqnarray}}

\newcommand{\la}{\left\langle}
\newcommand{\ra}{\right\rangle}

\renewcommand{\Im}{{\rm \, Im\,}}

\definecolor{nvgreen}{rgb}{0.0, 0.5, 0.2}

\begin{document}
	\sloppy
	
	\title{Multi-mode correlations and the entropy of turbulence in shell models}
	\author{Gregory Falkovich$^{1,2}$, Yotam Kadish$^1$ and Natalia Vladimirova$^{2,3}$}
	\affiliation{$^{1}$Weizmann Institute of Science, Rehovot 76100 Israel\\
		$^{2}$Landau Institute for Theoretical Physics, 142432 Chernogolovka, Russia\\$^{3}$Brown University,  Providence, RI 02912, USA
		%\\\mbox{${^3}$University of New Mexico, Albuquerque, NM 87131, USA}
	}
	%\maketitle
	
	\date{\today}
	
	\begin{abstract}
		We suggest a new focus for turbulence studies ---  multi-mode correlations --- which reveal the hitherto hidden nature of turbulent state. We apply this approach to shell models describing basic properties of  turbulence. The family of such models allows one to study turbulence close to thermal equilibrium, which happens when the interaction time weakly depends  on the mode number. As the number of modes increases, the one-mode statistics  approaches  Gaussian  (like in weak turbulence), the occupation numbers grow, while the three-mode cumulant describing the energy flux stays constant. Yet we find that higher multi-mode cumulants grow with the order. We derive analytically and confirm numerically the scaling law of such growth. The sum of all squared dimensionless cumulants is equal to the relative entropy between the full multi-mode distribution and the Gaussian approximation of independent modes; we argue that the relative entropy could grow as the logarithm of the number of modes, similar to the %mutual information and
		entanglement entropy in  critical phenomena. Therefore, the multi-mode correlations give the new way to characterize turbulence states and possibly divide them into universality classes.
	\end{abstract}
	
	\maketitle
	
	%\vskip 0.1truecm

\section{Introduction}

        We define turbulence as a state where many degrees of freedom are deviated from thermal equilibrium. This usually happens when wavenumbers or frequencies of the modes excited are vastly different from those of the modes that dissipate. A statistically steady state is then a long cascade of excitations.
	
	Fluctuations in a wide interval of scales and times is the property shared by turbulence and critical phenomena. That analogy was discussed either in terms of power-law behavior of correlation functions \cite{MN,EG} or in terms of the probability distribution of  macroscopic quantities \cite{UNI}. However, the analogy was not advanced to reach deeper understanding of turbulence, in particular, reproduce  the  spectacular success of statistical physics in defining the universality classes of critical phenomena.
	
	We suggest to focus on the entropic characteristics of turbulence, following the approach developed for critical phenomena by Wilczek, Kitaev, Cardy and others \cite{log1,log2,log3}. While the  entropy itself is usually linear in the number of degrees of freedom, the quantum entanglement entropy or classical mutual information between different parts of the system or between a system and its environment could be logarithmic at criticality.   The factor in front of the logarithm is the central charge of the conformal field theory, which determines the universality class of the transition \cite{log1}. That approach to critical phenomena brought most progress in two dimensions, here we shall apply it to one-dimensional model of turbulence. 	
	
	We believe that  an entropic approach is also natural for turbulence, which must have much lower entropy than thermal equilibrium (at the same energy). How does the entropy deficit depend on the number of modes deviated from equilibrium or, in other terms, on the Reynolds number?
	%After the trivial extensive part of the entropy deficit related to distortion of equipartition is subtracted (or amplitudes normalized by their RMS values), one is left with correlations responsible for the cascade.
	Since the entropy deficit is information, where it is encoded?
	
	Two natural possibilities exist here:  correlations between points in space or between modes in Fourier space.
	The multi-point spatial correlations were first discovered in the simplest case of non-equilibrium imposed by a spatial gradient (of temperature, velocity, etc). So-called Dorfman-Cohen anomalies manifest themselves as infrared divergencies in the density expansion of kinetic coefficients (thermal conductivity, viscosity, diffusivity) \cite{DC2,DC3,DC4}. However, turbulence deviates from equilibrium by forcing some modes and dissipating other,  imposing different conditions in Fourier space rather than in real space. Therefore, the cascade nature of turbulence must manifest itself in inter-mode correlations.
	
	Since the very existence of a cascade hinges on interaction between modes, the traditional approach is usually focused on the lowest two nonzero moments: the occupation numbers and the flux \cite{Kol,ZLF,FV1,FouxonOz,Kritsuk,Galtier}, which is typically a three- or four-mode correlation function. % (when the cascade is of non-quadratic invariant, the flux is a higher moment \cite{FouxonOz,Kritsuk,Galtier}).
	Yet it was argued recently from an entropic analysis that even for a weak wave turbulence two moments are not sufficient to describe the statistics of a multi-mode system, which could be far from Gaussian~\cite{SF}.

	We suggest that the nature of a turbulent state is reflected in multi-mode correlations which can be quantified by the mutual information between  upscale and downscale parts of the cascade or between all modes. We hope that the mutual information can play here the role that the entanglement entropy plays in critical phenomena. 	
	For the formidable task of analyzing the multi-mode correlations, it is natural to start from the class of systems where they are weak and can be treated perturbatively.
	We find one such class where occupation numbers are close to thermal equipartition,  the single-mode statistics is close to Gaussian, and the (dimensionless) flux is small. We demonstrate that nonzero multi-mode dimensionless cumulants can be  all of the same order in such systems. When this is the case,  the mutual information  must grow as a logarithm of the total number of modes in the cascade, thus establishing a direct quantifiable link with classical and quantum critical phenomena.
	
	Developed turbulence with the same scaling as in equilibrium occurs in many  systems. One class is both direct and inverse cascades in the universal Nonlinear Schrodinger model in two dimensions \cite{FV1}, which describes cold atoms and all spectrally narrow wave spectra. Another example is a joint turbulence of interacting high and low-frequency waves abundant in geophysics and plasma physics~\cite{ZLF,MRZ}.

\section{Definition of the models}

We consider a very wide class of systems with strong interaction, described by the  equations with quadratic nonlinearity,
	\be
	i \dot a_j =\sum\nolimits_{sq}\bigl(2V^q_{js} a_s^{*}a_q + V^j_{sq} a_s a_q \bigr)\,,
	\label{Hw}
	\ee
conserving a quadratic integral of motion, $E=\sum \omega_j  |a_j|^2$, as long as $V^j_{sq} \not=0$  iff $\omega_s+\omega_q=\omega_j$.	
Such are 	the Euler equations for incompressible fluid flows and for solid body rotations, as well as the family of two-dimensional hydrodynamic models from geophysics, astrophysics and plasma physics,  where a scalar field $a$ (vorticity, temperature, potential) is transported by the velocity whose stream function $\psi$ is linearly related to $a$: $\partial a/\partial t=-({\bf v}\cdot\nabla)a$, ${\bf v}=(\partial\psi/\partial y,-\partial\psi/\partial x)$, $\psi({\bf r})=\int d{\bf r}'|{\bf r}-{\bf r}'|^{\beta-2}a({\bf r}')$. For the 2D Euler equation, $\beta=2$. Other cases include surface geostrophic ($\beta=1$)~\cite{Holton},  rotating shallow fluid or magnetized plasma ($\beta=-2$)~\cite{Hasegawa}, etc. After Fourier transform, $\dot a_{\bf k} =\sum\nolimits_{\bf q} q^{-\beta}[{\bf k}\times{\bf q}]a_{\bf q}a_{\bf k-q}$.
	
Recently, the family was expanded further by adding models describing waves or modes with a resonant triplet interaction determined by the Hamiltonian \cite{Fibo,SVF}:
	\[ {\cal H}_w =\sum\nolimits_j\omega_j|a_j|^2+\sum\nolimits_{j,s,q}  V^j_{sq}\left( a_q^{*}a_{s}^{*}a_{j}+ a_sa_{q}a_{j}^{*}    \right). \]
	%\label{genA}
	Gauge transformation, $a_j\mapsto a_j\exp(i \omega_j t)$,
	turns  $i \dot a_j={\partial{\cal H}_w/ \partial a_j^*}$ into  (\ref{Hw}).

All models are scale-invariant: $V(\lambda \omega_j,\lambda \omega_s,\lambda\omega_q)=\lambda^\alpha V(\omega_j, \omega_s,\omega_q)$; for hydrodynamic models $\alpha=2-\beta$.
	They all have exactly Gaussian statistics in thermal equilibrium \cite{Fibo,SVF}: $\ln{\cal P_G}\propto -E/T$ and $\langle |a_j|^{2m}\rangle=\Gamma(m+1)n_j^{m}$, where $ n_j\equiv\la|a_j|^2\ra=T/\omega_j$. Thermal equipartition takes place both in a closed system and under a contact with a thermostat, which pumps and dissipates all $N$ modes equally, see  below. Turbulence appears when pumping and damping  act on distant modes, $p$ and $d$ respectively.  Nonlinear interaction provides for a cascade of excitations via the transparency window between $p$ and $d$.
	Cascade is called direct for $p=1, d=N$, and inverse for $p=N, d=1$. In the transparency window,
	\begin{align}
	\omega_{j}{\text{d}\langle |a_j|^2\rangle/ \text{d}t}=\Pi_{j-1}-\Pi_j \,,
	\label{gen2}
	\end{align}
	due to conservation of $E$. The right hand side is the discrete divergence of the flux
	$\Pi_j\simeq \sum_{s,q}V^j_{sq}\omega_j\la C_3(j,s,q)\ra$, where $C_3(j,s,q)=\Im\{a^*_{j}a_{s}a_q\}$ and the brackets $\la\cdot\ra$ denote time averaging. We also denote by $C_m$ with $m=3,4,\ldots,N$ generic multiplications of $m$ modes, for which the averages are the correlation functions with no reducible parts (zero for Gaussian statistics).  For scale-invariant systems, stationarity requires the flux independence on $j$, which sets the scaling $a_j^3\propto \Pi\omega_j^{-1-\alpha}$. Strong turbulence of resonantly interacting waves is different from weak turbulence determined by quasi-resonances, where $a_j^4\propto \Pi$~\cite{ZLF}.

	The rest of the article is devoted to the case of $\alpha=1/2$, as it allows us to explore the difference between statistics of turbulence and of thermal equilibrium.  One might expect a minimal difference, since the scaling of turbulence and equipartition coincide,  $a_j\propto \omega_j^{-(1+\alpha)/3}= \omega_j^{-1/2}$.  Yet, turbulence carries the flux from excess to scarcity, so the energy density $E_k=\omega_kn_k$ is not constant but must decrease from pumping to dissipation. This immediately brings the higher cumulants, beyond $C_3$. Indeed, $dC_3/dt$ is expressed via the fourth-order moments, whose cumulants are denoted collectively $C_4$. In a steady state,

	\small
	\bea
	&{\text{d}\la C_3(j,s,q)\ra / \text{d}t}=\sum_{kl}\bigl(V_{kl}^q\langle a_j^*a_ka_la_s\rangle+V_{kl}^s\langle a_j^*a_ka_la_q\rangle\nonumber\\& +2V_{ls}^k\langle a_j^*a_ka_l^*a_q\rangle+2V_{lq}^k\langle a_j^*a_ka_l^*a_s\rangle -V_{kl}^j\langle a_l^*a_k^*a_sa_q\rangle
	\label{c4}\\&\!-V_{jl}^k\langle a_k^*a_la_sa_q\rangle\bigr)\!=\! \la C_4\ra+2V_{sq}^j(n_sn_q-n_jn_s-n_jn_q)\!=\!0\,.\nonumber\eea
	\normalsize

\noindent
	In equilibrium, $n_k\propto\omega_k^{-1}$  and $\la C_4\ra = -2V_{sq}^j(n_sn_q-n_jn_s-n_jn_q) \propto -V_{sq}^j(\omega_j-\omega_q-\omega_s)=0$.
	The scaling of turbulence is the same, so the decrease of $\omega_kn_k$ towards the damping must be  logarithmic \cite{ZLF}. For a direct cascade, we denote $ \Lambda_k \equiv\log(\omega_d/\omega_k)$ and assume $\omega_k n_k=   \Lambda_k^\xi$, with $\xi$ to be determined. Considering $j-s\simeq j-q\ll j$ we expand $n_s,n_q$ around $n_j$   (see the Supplement for the details), which gives:  $\la C_4(j)\ra\approx 2\xi V_{sq}^j n_j^2/\Lambda_j\propto \Lambda_j^{2\xi-1}$ for $\Lambda_j\gg1$. Stationarity of $C_4,C_5$ give respectively $\la C_5\ra\simeq Vn\la C_3\ra$, $\la C_6\ra\simeq Vn\la C_4\ra\simeq Vn^3/\Lambda$, and generally $\la C_m(j)\ra\simeq Vn_j^{m/2}/\Lambda\propto\omega_j^{-m/2}\Lambda^{\xi m/2-1}$. The flux law requires that $C_3$ has no logarithm, which gives $\xi=2/3$. We thus see that the dimensionless cumulants of all orders  are suppressed by the same factor:  $\la C_m(j)\ra n_j^{-m/2}\simeq \log^{-1}(\omega_d/\omega_j)$. Cumulants are of order unity at $j\simeq d$ \cite{Fibo}.  As $d=N$ increases, the cumulants at $j\ll d$ are getting uniformly smaller but their number increases, since the cumulants of  orders up to $m\simeq d-j$ involve the   mode $j$.
	
\section{Computation of cumulants}

        We now present a direct computation for two models, defined respectively by the Hamiltonians \cite{Fibo,SVF}:
	\begin{align}
	&{\cal H}_2=\sum\nolimits_j2^{j\alpha}(a_j^2a_{j+1}^*+{a_j^*}^2a_{j+1})\,,\label{FibA}\\
	&{\cal H}_3 =\sum\nolimits_{j}  F_j^\alpha\left( a_j^{*}a_{j+1}^{*}a_{j+2}+  a_ja_{j+1}a_{j+2}^{*}\right)\,.\label{FibB}
	\end{align}
	The first (doubling) model describes a chain of modes resonantly interacting with their second harmonics. The second (Fibonacci) model describes a chain of resonantly interacting triplets; here  $F_j$ are the Fibonacci numbers, defined by $F_j+F_{j+1}=F_{j+2}$.
	%	The Hamiltonians are invariant with respect to the gauge transformation $a_j\to a_j e^{i \omega_jt}$ which provide conservation of  $E=\sum _{j} \omega_{j }|a_j|^2$, where  respectively  $\omega_j=2^j$ or $\omega_j=F_j$.
	The Hamiltonians are invariant with respect to the gauge transformation with $\omega_j=2^j$ and $\omega_j=F_j$ respectively.
	Since  $F_j=[\phi^j-(-\phi)^{-j}]/\sqrt5$, then at $j\gg1$, the wave frequency also depends exponentially on the mode number,  $F_j\propto \phi^{j}$, where $\phi=(1+\sqrt5)/2$ is the golden mean.  Both (\ref{FibA},\ref{FibB}) are so-called shell models defined in logarithmically discretized Fourier space \cite{Fibo,SVF,Bif,Pro}.
	
	%We add dissipation and white-in-time pumping:
	%$\dot a_j=-i\partial {\cal H}/\partial a_j^*+\eta_j-\gamma_ja_j$.
	%Here $\langle \eta_ja_k^*\rangle=\delta_{jk}P_j/2$, and $P_j$ is the power of pumping. If pumping and damping provide equipartition, so that $P_j \omega_{j}/\gamma_j$ is independent of $j$, the thermal equilibrium distribution  is Gaussian, $\ln{\cal P}\propto - E $, and $\langle |a_j|^2\rangle=P_j/ {2}\gamma_j\propto \omega_j^{-1}$ \cite{Fibo}.

	To study turbulence, we pass to  $b_j=a_j\omega_j^{ (1+\alpha)/3}$, which makes the triple moment $j$-independent in the transparency window. We also renormalize time to bring the dynamical equations to the universal form (see numerical factors in the Supplement): \begin{align}&i  \dot b_j
	= 2^{j{2\alpha-1\over3}} \left( b_{j-1}^2  + b_{j}^{*}b_{j+1 }\right)\,.\label{doub}\\&
	\!\!\!
	i  \dot b_j
	= \phi^{j{2\alpha-1\over3}} [\phi b_{j-1} b_{j-2}\!+ b_{j-1}^{*}b_{j+1}\!+\phi^{-1}b_{j+1}^{*}b_{j+2}]\,.\label{Fibb}
	\end{align}
	Remarkably, both equations have an exact stationary solution $b_j=\imath$, $\forall j$. It corresponds to the turbulent scaling $a_j \propto \omega_j^{-(1+\alpha)/3}$. The solution is linearly unstable and cannot be matched with pumping and damping,  yet we find that turbulence fluctuates around it. The prefactors $\omega_j^{(2\alpha-1)/3}$ ($=2^{j(2\alpha-1)/3}$, $\phi^{j(2\alpha-1)/3}$) determine the dependence of the interaction time on the mode number $j$.
	For $\alpha=1/2$, the equations (\ref{doub},\ref{Fibb}) are translation invariant along $j$, the interaction time is the same for all modes, and scaling laws of turbulence and equilibrium coincide. Both direct and inverse cascades exist in this case~\cite{Fibo,SVF}.

	Analytic derivations of cumulants for (\ref{FibA}) are presented here and for (\ref{FibB}) in the Supplement. Consider the products
	$C_{kl\bar m\ldots}(j)=  b_{j-k}b_{j-l} b_{j-m}^*\ldots b_{j}^*$, having non-zero mean values when gauge invariant:  $2^{-k}+2^{-l}-2^{-m}+\ldots-1=0$.
	Every such product of order $m$  can be obtained from that of $m-1$  replacing $b_j$ by either $b_{j-1}^2$ or $b_{j}^*b_{j+1}$.
	%For example,  $C_4= C_{221\bar0}(j)=  b_{j-2}^2 b_{j-1}b_{j}^* $, and $C_{3321\bar0}(j)=  b_{j-3}^2b_{j-2} b_{j-1}b_{j}^* $,   $C_{2222\bar0}(j)=  b_{j-2}^4b_{j}^* $ are both $C_5$.
	Denote $A_m(j) =\langle |b_j|^{m}\rangle$. Stationarity of $A_2$  similar to (\ref{gen2}) gives $\langle C_3 \rangle =\langle C_{11\bar0}(j) \rangle =1$. Next,
	\bea &d\langle b_j^2b_{j+1}^*\rangle/dt= 2\langle|b_jb_{j+1}|^2\rangle-\langle|b_j|^4\rangle+2\langle C_{221\bar0}(j+1)\rangle\nonumber\\&\!\!-\langle C_{221\bar0}(j+2)\rangle = 2n_j(n_{j+1}-n_j)+\langle\langle C_4(j)\rangle\rangle=0\,,\label{double}\eea
	where we denoted the combination of the fourth-order cumulants: $\langle\langle C_4\rangle\rangle =\langle C_{221\bar0}\rangle+\langle\langle2|b_jb_{j+1}|^2- |b_j|^4\rangle\rangle$.	 Double brackets denote subtraction of the reducible parts.  Assuming self-consistently that the statistics is close to Gaussian and $n_j\propto \log^\xi (2^{j-d})\propto |j-d|^{\xi}$, we obtain  $\langle\langle C_4(j)\rangle\rangle\approx   2n_j(n_j -n_{j+1})\approx 2\xi n_j^2/|j-d|$. Numerics described below confirm that.   Next, stationarity of $\langle |b_j|^4\rangle$ gives $ \langle |b_j|^2C_3(j)\rangle= \langle |b_j|^2C_3(j+1)\rangle$. Since the reducible contributions are respectively 3 and 2 times $n_j\langle C_3\rangle$, then the difference of the cumulants is  $\langle\langle |b_j|^2[C_3(j) -C_3(j+1)]\rangle\rangle=A_2(j)\langle C_3\rangle=n_j$, that is the fifth-order cumulants are comparable to $n_j$. Numerics  give   $ \langle |b_j|^2C_3(j)\rangle= \langle |b_j|^2C_3(j+1)\rangle \approx 2.25n_j$. Stationarity of $\langle |b_jb_{j+k}^*|^2\rangle$ by induction over $k$ imposes the identities: $ \langle |b_j|^2[C_3(j+k+1)-C_3(j-k)]\rangle=0$, the cumulants are negligible for $k>1$ when there is no overlap (see Supplement).
	Next orders give  $\langle\langle C_6\rangle\rangle \simeq A_2\langle\langle C_4\rangle\rangle \simeq A_2A_4/|j-d|\simeq A_6/|j-d|$  and  $\langle\langle C_m\rangle\rangle \simeq A_m/|j-d|$, supporting the above estimates in the general case.
	Since $A_m\propto n_j^{m/2}$, then $\langle\langle C_m\rangle\rangle =n_j^{m/2}/|j-d|\propto |j-d|^{m\xi-1}$. Applying it to $\langle C_3\rangle =1$ we obtain $\xi=2/3$, that is $n_j\propto |j-d|^{2/3}$,
	$A_m(j)\simeq n_j^{m/2}\propto |j-d|^{m/3}$ and $\langle\langle C_m(j)\rangle\rangle\simeq {A_m(j)/|j-d|}\propto |j-d|^{(m-3)/3}$. We see that cumulants with $m>3$ actually grow with increasing the length of the cascade.
	The dimensionless cumulants $D_m=\langle\langle C_m\rangle\rangle/\langle |C_m^2|\rangle^{1/2}$ all decay by the same law:
	\be D_{kl\bar s\ldots}(j)\equiv{\langle \langle b_{j-k}b_{j-l} b_{j-s}^*\ldots b_{j}^*\rangle\rangle\over\langle  |b_{j-k}b_{j-l} b_{j-s}^*\ldots b_{j}^*|^2\rangle^{1/2}}\simeq {1\over|j-d|}\,.\label{Dijk}\ee

	\section {Numerical results}

\begin{figure}[h!]
		\centering
		\includegraphics[width=85mm]{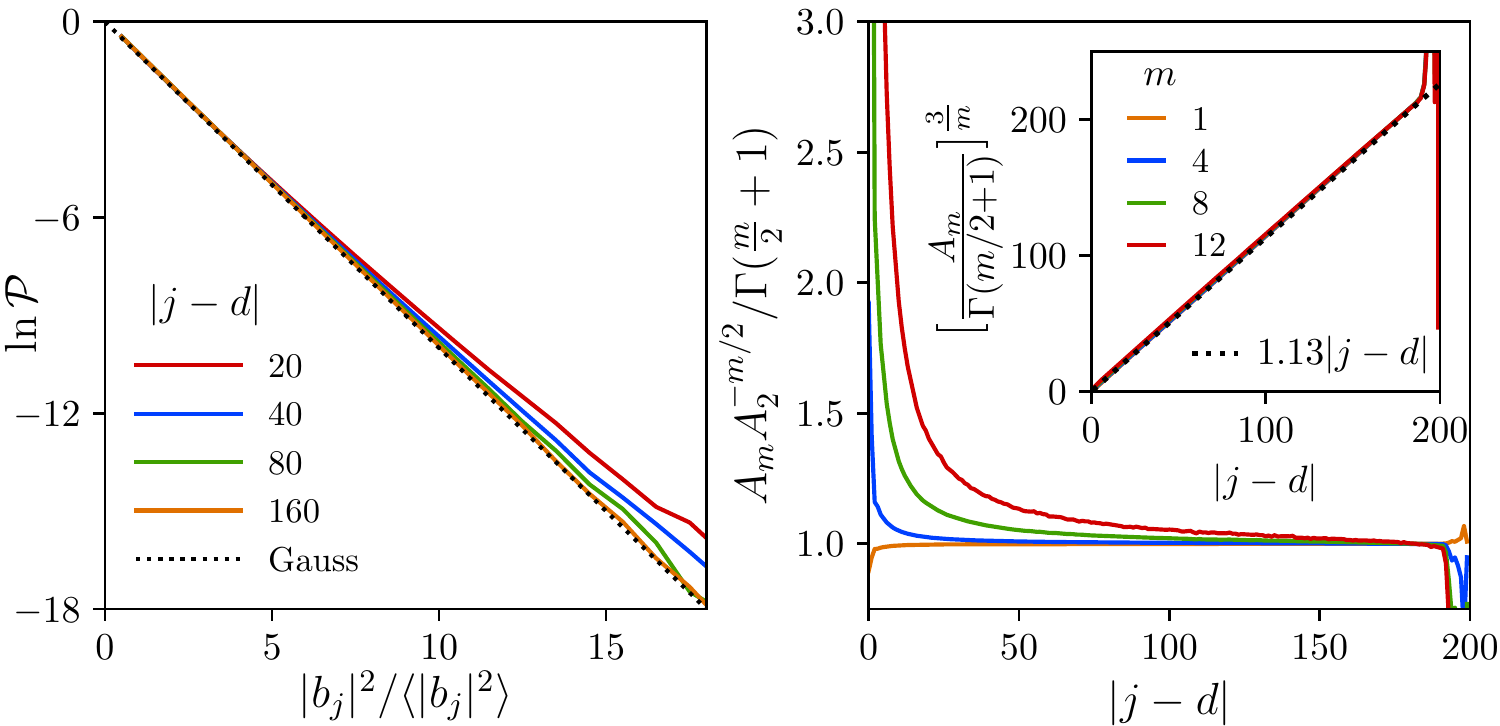}
		\caption{ Direct Fibonacci cascade. Left: away from damping, single-mode distribution approaches
			Gaussian. Right: single-mode moments
			$A_m=\langle |b_j|^{m}\rangle$. }
		\label{probN}
		\label{An}
	\end{figure}

        To check the assumptions in our derivations, we solved numerically   (\ref{doub},\ref{Fibb}) for $\alpha=1/2$ and both direct and inverse cascades. Exponentially large $\omega_i$ do not appear  for the variables $b_i$, which allows us to enlarge $N$ without shortening the time step. Compared with ~\cite{Bif,Fibo,SVF}, this novel approach gives much larger transparency window, up to a record $N=200$. Figure~\ref{An} shows that the single-mode probability distribution is close to Gaussian for all modes in the transparency window, see also \cite{SVF,Fibo}.
	
	It also shows the dependence of the single-mode moments $A_m(j)$  on $j$ counted from dissipation. The occupation numbers are of order unity at $j\simeq d$; away from dissipation, the moments are indeed Gaussian, that is the renormalized moments $A_m(j)A_2^{-m/2}(j)/\Gamma(1+m/2)$ are all equal to unity, independent of $j$. That means that the single-mode distribution is scale invariant, the scaling is indeed $A_m(j)\propto |j-d|^{m/3}$. The same is true for the doubling model.

\begin{figure}[h!]
		\centering
		\includegraphics[width=87mm]{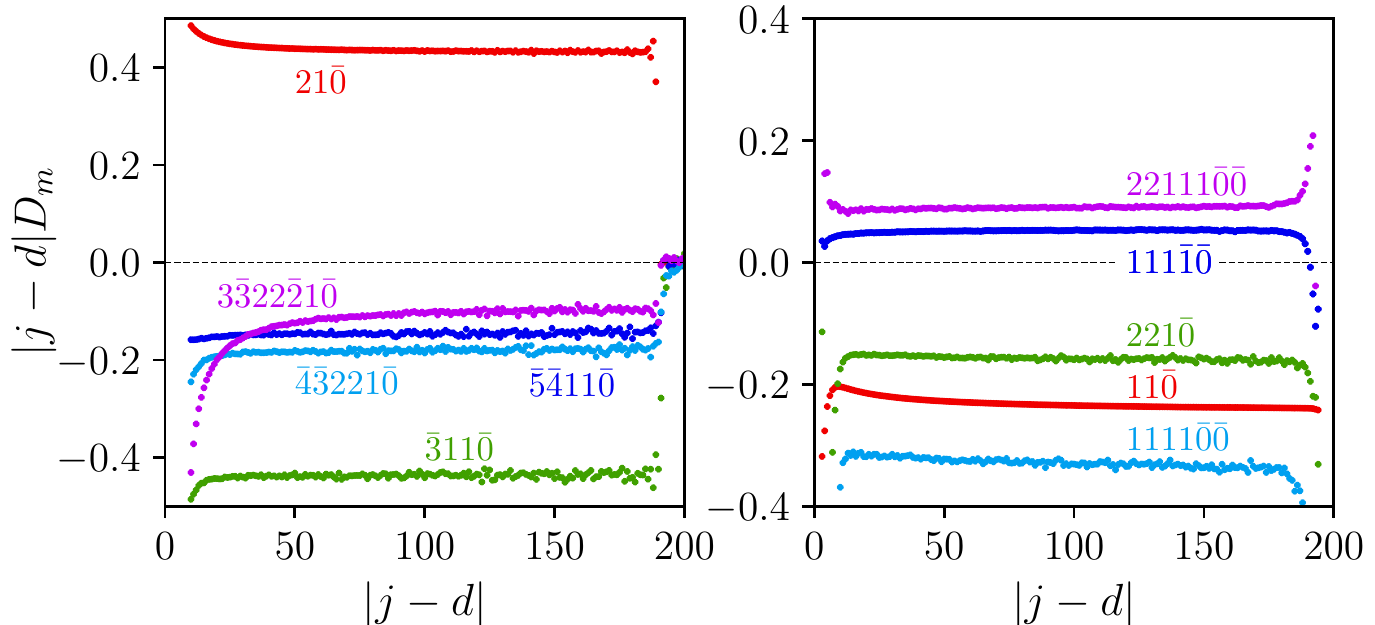}
		\caption{Largest compensated cumulants for every order. Left: Direct cascade in the Fibonacci model.
			Right: Inverse cascade in the doubling model.}
		\label{fiboflat}
	\end{figure}

	Both large transparency window and massive statistics are crucial for the first-ever computation of high-order cumulants in turbulence presented in
	Figure~\ref{fiboflat}. Due to symmetry $b_j\to -b_j^*$, the cumulants with even $m$ are real and those with the odd $m$ are imaginary (they are non-zero due the breakdown of time reversibility in turbulence). The cumulants of $b_j'=i b_j$  are all real and proportional to the flux, that is have opposite signs for direct and inverse cascades (see also Supplement). Figure~\ref{fiboflat} shows dimensionless cumulants multiplied by $|j-d|$.
	The data confirm the scaling   (\ref{Dijk}) for all pure cumulants, like $\langle C_{221\bar0}\rangle$, and those comparable to the reducible part, like $ \langle |b_j|^2C_3(j)\rangle$. The  few cumulants due to amplitude-only correlations, like $\langle\langle |b_jb_{j-1}|^2\rangle\rangle$,  are small differences of large values; the data are irregular, don't contribute the mutual information, and left out of the consideration.
	%It could be that on top of $|j-d|^{-1}$ dependence, there could be even weaker dependence on the interval width (see Supplement).
	
	Numerics  give the squared flux  $\langle C_3^2\rangle= \langle(b_{j-1}^2b_j^*)^2\rangle \approx 2.44n_j^3/|j-d|$ for (\ref{FibA}).
	Two things are noteworthy. First, $ \langle C_3^2\rangle\propto |j-d| \gg  \langle C_3\rangle^2=1$, that is the flux fluctuates strongly, its variance grows unbounded with $N$, while its mean stays constant. The same is true for (\ref{FibB}), and reproduces the experimental and numerical data on direct and inverse cascades in fluid turbulence \cite{FC}. Second, for both models, we find no systematic decay of cumulants with $m$. We cannot presently prove this for $m\to\infty$, nor we are able to classify all gauge-invariant cumulants of an arbitrary order. We managed it up to the seventh order: (\ref{FibA}) gives 26 distinct types of cumulants and  (\ref{FibB}) gives 70 for $m=7$  (see the tables in the Supplement). For example, cumulants of the type $\langle\langle C_3C_4\rangle\rangle$ are among the largest and are  comparable to $\langle C_4\rangle \langle C_3\rangle=\langle C_4\rangle$ --- another sign of  no decay with $m$ (see also the Table).

	To summarize the findings: the dimensionless cumulants are uniformly small far from dissipation: $D_m(j)\simeq |j-d|^{-1}$. In the thermodynamic limit $d=N\to\infty$ (direct cascade), all dimensionless cumulants tend to zero and we have an asymptotic equipartition with the temperature $T=N^{2/3}$ for all finite $j$.  Yet the number of cumulants grow with $N$, so it is not clear if the full multi-mode statistics  approaches Gaussian in the limit.
	
\section{Entropic consideration}

        Let us  argue that the growth of cumulants makes the full multi-mode probability distribution  ${\cal P}\{b_j\}$ very different from the Gaussian distribution of independent modes with the same occupation numbers, ${\cal  P}_G\{b_j\} =\prod\nolimits_jn_j^{-1}\exp\left(- |b_j|^2/n_j\right)$.
	Since  the dimensionless cumulants are small, the probability distribution can
	be approximated as follows:
	\bea &{\cal P}\{b_j\}=Z^{-1}\exp\left[-\sum\nolimits_{j=1}^N\alpha_j|b_j|^2+B\right]\,,\label{P}\\
	&Z\approx (1+\langle B^2\rangle_0/2)\prod\nolimits_j\alpha_j^{-1}\,. \label{Zn} \eea
	The average $\langle \ldots\rangle_0$ is with ${\cal P}_G$, and $B$ is the sum of all products giving nonzero cumulants, symbolically
	\be B=\sum\nolimits_m{C_m\langle C_m^*\rangle+C_m^*\langle C_m\rangle\over\langle |C_m|^2\rangle}\,.\label{sum}\ee
	Define $S_N=\la\ln{1/\cal P}\ra$ as the total entropy. Relative entropy, ${\cal D}({\cal P}|{\cal P}_G)=\langle \ln({\cal P}/{\cal P}_G)\rangle$ measures the difference between ${\cal P}$ and ${\cal P}_G$. Since the latter is a product, ${\cal P}_G\propto \prod\nolimits_j\exp(-\alpha_j|b_j|^2)$, then ${\cal D}=\sum_{j=1}^NS_j -S_N$ is also the difference of the entropies, that is the multi-mode mutual information $I_N$, quantifying correlations in the system.
	
	In the leading order in $1/|j-d|$, ${\cal D}$ is the sum of all squared dimensionless cumulants.
	If we had only triple cumulant,   then ${\cal D}$ would be given by a converging series and thus independent of the number of modes:
	{ ${\cal D}=\sum_{j=3}^{N}{  D_{3} ^2 (j)/4} \simeq \sum_{j=3}^{N} (N-j)^{-2}\simeq 1$.}
	The contributions of multi-mode cumulants makes  ${\cal D}$ a double sum.  Our computations suggest that the sum of squared dimensionless cumulants do not decrease with the order: Table~\ref{table} shows quite irregular dependence  on $m$, but no overall decay: contributions of $m=4$ is comparable to $m=6$, while $m=5$  to that of $m=7$. This is all the more remarkable since only nearest neighbors interact. Assuming that indeed the sum of squared cumulants do not decay with the order and asymptotically saturates to some number (which we cannot yet compute) $c=\lim_{m\to\infty}|  D_m(j) |^2|j-d|^2/4$, we obtain the relative entropy  logarithmic in the number of modes:
	\be
	{\cal D}= \sum_{m=3}^{N}\sum_{j=m}^{N} {| D_m(j) |^2\over 4} = c\sum_{m=3}^{N}\sum_{j=m}^{N}  {1\over j^{2}}=c\log N\,.\label{RE}\ee
	Since we disregard amplitude-only correlations, the conclusion is about ${\cal D}({\cal P}\{b_j\}|{\cal P}\{|b_j|\})$ which bounds  ${\cal D}({\cal P}|{\cal P}_G)$ from below. The mutual information between the two parts of the cascade, $I(A,B)=S(A)+S(B)-S(A,B)$, must have the same logarithmic dependence, since it is
	given by the double sum over $j,m$ of the squared cumulants that involve modes from both $A=\{p\ldots N/2\}$ and  $B=\{1+ N/2\ldots d\}$.
	The longer is the transparency window, the smaller are the cumulants, yet larger is their number, so that the mutual information grows logarithmically with the number of modes excited, that is with the effective Reynolds number. The reason for the growth here is the nonlocality of correlations. Another source of a logarithmic dependence of the mutual information on the degree of  non-equilibrium was found for a minimal (two-mode) model of a cascade \cite{min}. It is a task for the future to derive the logarithmic law with the factor $c$ for different models, which requires solving a formidable problem of classification of cumulants of an arbitrary order. After that, we may be able distinguish universality classes of turbulent cascades.
	
	\section{Discussion}

        The next step will be to study multi-mode correlations  for  $\alpha\not=1/2$, when non-Gaussianity of a single mode grows exponentially along the cascade: $A_{2m}/A_2^m\propto \omega_{j-p}^{\Delta_{2m}}$ \cite{Fibo,Bif,SVF}. Preliminary (short-interval) data show a linear growth with $|j-p|$ for $I(b_{j-1},b_j)$ and $I(b_{j-1},b_j,b_{j+1})$ \cite{Fibo,SVF}, which bound $I(b_1,\ldots,b_N)$  from below due to monotonicity. It is tempting to treat $\alpha=1/2$ logarithmic case as that of critical phenomena in  dimension 4 \cite{LK} and develop a  Wilson-type $\epsilon$-expansion in $\epsilon=\alpha-1/2$ \cite{Wilson}. That could be non-trivial, since the perturbation expansions, regular in thermal equilibrium, tend to be singular in non-equilibrium states, especially in turbulence \cite{SF,DC2,DC3}.

	\begin{table}
		\begin{tabular}{ ccccc}
			\hline
			\makebox[6mm]{$m$}
			& \makebox[16mm]{${\cal H}_3$ direct} &  \makebox[16mm]{${\cal H}_3$ inverse}
			&  \makebox[16mm]{${\cal H}_2$ direct} & \makebox[16mm]{${\cal H}_2$ inverse}\\
			\hline
			3 & 0.17986 & 0.18992 & 0.06522 & 0.05660\\
			4 & 0.22636 & 0.22748 & 0.02818 & 0.02558 \\
			5 & 0.07100 & 0.07102 & 0.00600 & 0.00356\\
			6 & 0.11829 & 0.12089 & 0.19594 & 0.17387\\
			7 & 0.06500 & 0.06670 & 0.02553 & 0.02333\\
			\hline
			
		\end{tabular}
		\caption{ Sum of squared dimensionless cumulants,  $D_m^2(j)|j-d|^2$, in the transparency window for both models.
			%\\ These cumulants were computed as follows: The irreducible part is a subtraction between the cumulant and all his possible representations as a multiplication between lower order cumulants (up to first order expansion of \ref{P}). A division with the Gaussian rms made the result dimensionless. To get a constant value over the inertial range, the squared irreducible-dimensionless cumulant was multiplied by $|j-d|^2$.
		}\label{table}
	\end{table}
	
	Note the dramatic difference between  our turbulence and  nonuniform dilute gases described in \cite{DC2,DC3,DC4}. There, even though naive expansion encounters divergencies, renormalized expansion  gives  higher cumulants proportional to higher powers of the small parameter (density), which leaves the multi-particle mutual information small. Our (\ref{Dijk}) gives all cumulants proportional to the same (first) power of the small parameter, which may lead to logarithmically large multi-mode mutual information.	
	
	Our models deal only with resonant modes and respective cumulants. In a general wave turbulence, both resonant and non-resonant interactions are present, so the analysis of multi-mode correlations will be more complicated. It was argued in \cite{SF} that cumulants might be substantial for resonant modes in the weak turbulence, even when the statistics of mode amplitudes is close to Gaussian. It is likely that in a continuous limit, determined by quasi-resonances rather than resonances, the higher cumulants are proportional to higher powers of the small parameter, and the mutual information is small.
	As far as real-world fluid turbulence is concerned, we believe that our work shows importance of measuring multi-mode correlations and the entropy of multi-mode distributions. While only treatment of a moderate number of modes is feasible, it may provide an important insight.
	
	\section{Conclusion}
        We have shown that turbulent cascade necessary involves multi-mode correlations even in systems with local interaction and close to equilibrium. The entropy-lowering information about turbulence is encoded in the overlapping sets of multi-mode correlations. As far as we were able to compute, the degree of correlation does not decay with the number of modes. If true, multi-mode cumulants would make the multi-mode statistics very different from Gaussian; the logarithmic growth of the relative entropy with the number of modes establishes a promising analogy with critical phenomena. 	It remains to be seen how universal are multi-mode correlations across classes of turbulent systems and whether they can be related to coherent structures.

	We thank A. Zamolodchikov and M. Shavit for useful discussions. GF is grateful to \fbox{K. Gawedzki} for inspiration and to NYU and Simons Center for their hospitality. The work was supported by the Excellence Center at WIS, and by the grants  662962 and 617006 of the Simons  Foundation,  075-15-2022-1099 of the Russian Ministry of Science and Higher Educations, 823937 and 873028 of the EU Horizon 2020 programme, and 2018033 and 2020765 of the BSF.  NV was in part supported by NSF grant number DMS-1814619. This work used the Extreme Science and Engineering Discovery Environment (XSEDE), supported by NSF grant number ACI-1548562, allocation DMS-140028.

\end{document}

\sloppy
	
	\title{Supplement: Multi-mode correlations and the entropy of turbulence}
	\author{Gregory Falkovich$^{1,2}$, Yotam Kadish$^1$ and Natalia Vladimirova$^{2,3}$}
	\affiliation{$^{1}$Weizmann Institute of Science, Rehovot 76100 Israel\\
		$^{2}$Landau Institute for Theoretical Physics, 142432 Chernogolovka, Russia\\$^{3}$Brown University,  Providence, RI 02912, USA
		%\\\mbox{${^3}$University of New Mexico, Albuquerque, NM 87131, USA}
	}
	
	\date{\today}
	%\maketitle
	
	\begin{abstract}
		Here we list some technical details
		pertaining to analytic and numerical computations of
		multi-mode correlations in doubling and Fibonacci models
		defined in the main text.
	\end{abstract}
	
	\maketitle
	
	%--------------------------------------------------------
	\section{Derivation of higher order cumulants in a general triplet cascade}
	%--------------------------------------------------------
	
	Consider the three-wave Hamiltonian:
	\be
	\HH_w=\sum_{j,s,q}V^j_{sq}\lp a_s^*a_q^*a_j+a_sa_qa_j^*\rp
	\label{eqn: 3-wave Hamiltonian}
	\ee
	which admits the following dynamical equations:
	\be
	i\dot{a}_j=\sum_{s,q}\lp2V^s_{jq}a_q^*a_s+V^j_{sq}a_sa_q\rp
	\label{eqn: 3-wave eom}
	\ee
	
	\subsection{Expression for the flux}
	The energy flux through site $j$ is the rate of change of the total energy in sites $\ell\leq j$ (opposite direction for the inverse cascade). We compute this quantity explicitly from (\ref{eqn: 3-wave eom}):
	\be
	\begin{aligned}
		\Pi_j=&-\sum_{\ell\leq  j}\fd{\omega_\ell\la|a_\ell|^2\ra}{t}=\\
		&-\sum_{\ell\leq j}\omega_\ell\la \dot{a}_\ell^*a_\ell+a_\ell^*\dot{a}_\ell\ra=\la \sum_{sq}i\lp 2\omega_\ell V^s_{\ell q} a_q a_s^{*} + \omega_\ell V^\ell _{sq} a_s^{*} a_q^{*}\rp a_\ell  + c.c\ra=\\
		&=-2\sum_{\ell\leq j}\sum_{sq} \omega_\ell \lb 2V^s_{\ell q} C_3(\ell ,q;s) -  V^\ell _{sq} C_3(s,q;\ell )\rb=\\
		&=-4\sum_{\ell\leq j}\sum_{s}\sum_{q\leq s} \omega_\ell \lb V^s_{\ell q} C_3(\ell ,q;s) -  V^\ell _{sq} C_3(s,q;\ell )\rb
	\end{aligned}
	\ee
	In the final row, the right expression is exactly all the terms of the form $V^\ell _{sq} C_3(s,q;\ell )$ such that $\ell\leq j$ and $\omega_s+\omega_q=\omega_\ell$. The left expression also contains these, but also the terms $V^s_{\ell q} C_3(\ell ,q;s)$ such that $s>j$, $\ell\leq j$ and $\omega_\ell+\omega_q=\omega_s$. So we are left exactly with those terms, i.e.:
	\be
	\Pi_j = -4\sum_{\ell\leq j}\sum_{s>j}\sum_q \omega_\ell V^s_{\ell q} C_3(\ell ,q;s)
	\ee
	We see that $\Pi_j-\Pi_{j-1}\propto\omega_j^{-1-\alpha}$.  If each site interacts with only a small number of relatively near-by sites, then this sum has only few terms, and the flux is close to $\Pi_j\simeq\omega_jV_jC_3(j)$.

	\subsection{Estimation of $C_4$}
	Under assumption of interaction locality, it is enough to consider interacting triplets only sites $j,s,q$ such that  $j-q\simeq j-s\ll j$ while writing the equation for the fourth order cumulants $\la C_4\ra$:
	{\small
		\be
		\begin{aligned}
			&\Im\fd{\la a_qa_sa_j^*\ra}{t}=\Im\la\dot{a}_qa_sa_j^*+a_q\dot{a}_sa_j^*+a_qa_s\dot{a}_j^*\ra\\
			&=\Re\la-\lp\sum_{kl}V^q_{kl}a_ka_l+2V^k_{ql}a_l^*a_k\rp a_sa_j^*-a_q\lp\sum_{kl}V^s_{kl}a_ka_l+2V^k_{sl}a_l^*a_k\rp a_j^*+a_qa_s\lp\sum_{kl}V^j_{kl}a_k^*a_l^*+2V^k_{jl}a_la_k^*\rp\ra\\
			&=\la C_4\ra+2V_{sq}^j\lb n_qn_s-n_jn_q-n_jn_s\rb
		\end{aligned}
		\ee
	}
	From stationarity this equation gives a steady state value for the fourth order cumulants in terms of differences between Gaussian moments which are non-zero in thermal equilibrium, but their differences are zero (this is a general feature of the dynamics, for cumulants of even orders). Stationarity gives:
	\be
	\begin{aligned}
		\la C_4 \ra&= -2V_{sq}^j\lb n_qn_s-n_jn_q-n_jn_s\rb
		=-2V_{sq}^j n_q n_s n_j\lb n_j^{-1}- n_s^{-1}- n_q^{-1} \rb{=}\\
		&=-2V_{sq}^j n_q n_s n_j\lb\omega_j\Lambda_j^{-\xi}-\omega_s\Lambda_s^{-\xi}-\omega_q\Lambda_q^{-\xi}\rb=-2V_{sq}^j n_q n_s n_j\lb\omega_j\Lambda_j^{-\xi}-\omega_s\Lambda_s^{-\xi}-(\omega_i-\omega_s)\Lambda_q^{-\xi}\rb\approx \\
		&= -2V_{sq}^j n_q n_s n_j \, \omega_j \xi \Lambda^{-\xi-1}_j  \left[ 1 - \frac{\omega^2_q + \omega^2_s}{\omega^2_j} \right]  = 4 V_{sq}^j n_q n_s \frac{\xi}{\Lambda}\frac{\omega_q \omega_s}{\omega^2_j} = 4 V_{sq}^j n^2_j \frac{\xi}{\Lambda_j} \frac{E_q E_s}{E^2_j}  \approx 4\xi V_{sq}^j n_j^2\Lambda_j^{-1}
	\end{aligned}
	\ee
	Here we substituted $\omega_kn_k=\log^{\xi}(\frac{\omega_d}{\omega_k})\equiv\Lambda_k^\xi$ and $\Lambda^{-\xi}_s \approx \Lambda^{-\xi}_j + \frac{\xi}{\omega_j}  \Lambda^{-\xi - 1}_j (\omega_s - \omega_j)$.

	%--------------------------------------------------------
	\section{Numerical setups and diagnostics}
	%--------------------------------------------------------
	
	\subsection{Fibonacci model}
	The flux, $\Pi_j\equiv -\sum\limits_{j}^m F_{j+k-1}{d\langle
		|a_j|^2\rangle\over dt} =
	2F_{j+1}V_{j-1}J_{j+1}+2F_{j}V_{j}J_{j+2}$, is expressed via the
	triple cumulant, $J_j \equiv \Im\langle a_j^*a_{j-1}
	a_{j-2}\rangle$. In a steady state, the flux is constant in the
	transparency window, where $P_j=\gamma_j=0$, which determines $ J_j=
	(PF_p/2)5^{\alpha/2}\phi^{-j(1+\alpha)+1+2\alpha}$.  The sign of the
	flux coincides with the sign of $2\alpha-1$; when $\alpha\not=1/2$ the
	cascade goes from slow to fast modes.  For the normalized time $t\to
	t\bigl(\phi^{1-\alpha}5^{-\alpha} PF_p/2\bigr)^{1/3}$ and amplitudes,
	$b_j=a_j\left[\phi^{j(1+\alpha)-1-2\alpha}2/PF_p\right]^{1/3}5^{-\alpha/6}$,
	the dynamic equation takes the form:
	\be
	{b}_j' = -i
	\left[  \phi\, b_{j-2}b_{j-1} +  b^*_{j-1} b_{j+1} + \phi^{-1} b^*_{j+1}b_{j+2} \right]
	\phi^{j(2\alpha - 1)/3} - \gamma_j b_j +\xi_j,
	\label{eqn: fibo chain}
	\ee
	The system above is evolved numerically using {\tt lsode} solver
	[A.C. Hindmarsh, ACM Signum Newsletter, \textbf{15}(4), 10, (1980),
	K. Radhakrishnan and A. C. Hindmarsh, the Livermore solver for ordinary differential equations (1993)]
	%\cite{lsode}.	
	The forcing is applied to a single mode, $p$. It is implemented
	to provide solutions independent of time step, $\Delta t$, in the
	limit of small $\Delta t$,
	\[
	\xi_p = \frac{\sqrt{P_p}}{\sqrt{\Delta t}} \left(  r_{re} + \imath \,r_{im} \right),
	\quad\quad{\rm with} \quad
	P_p = 2 \sqrt{5} \, \phi^{p(2\alpha - 1)/3 - (1+\alpha)},
	\]
	where $r_{re}$ and $r_{im}$ are random numbers taken from distribution
	$\; \mathcal{P} = \pi^{-1/2} \exp \left( -r^2 \right)$ with $\la
	r_{re}^2 \ra = \la r_{im}^2 \ra = \sqrt{2}/2$, so that $\la |r|^2\ra = 1$.
	Damping $\gamma = \gamma_d = \gamma_{d \pm 1}$ is applied only to the two
	modes at the end of interval where the flux is going, that is to modes
	1 and 2 for inverse cascades, and to modes 199 and 200 for the direct
	cascade. All simulations are done in the system of 200 modes with
	$\alpha=1/2$.  Pumping is applied to mode $p=10$ and to mode $p=190$ for
	the direct and inverse cascades respectively.  Note that we do not
	control the amplitude of pumping since it is absorbed into rescaled
	time. Thus, $\gamma=0.67$ is the only empirical parameter in
	simulations; it is selected to minimize the built-up in occupation
	numbers at the end of the interval.  Simulations are done at time step
	$\Delta t = 0.01$, while complex amplitudes of all modes are recorded every
	10th step.  We have collected $5\times 10^8$ snapshots over multiple
	realizations.

	\subsection{Doubling model}
	
	The frequency doubling Hamiltonian for $\alpha=\frac{1}{2}$ is
	\[
	\HH = \sum _{j=1}^N 2^{\frac{j}{2}} \lp a_j^{*2}a_{j+1}+a_j^2 a_{j+1}^*\rp\ ,
	\]
	hence the Hamiltonian equations of motion are $\Dot{a}_j=-i 2^{1+\frac{j}{2}} a_j^* a_{j+1} -i2^{\frac{j-1}{2}}a_{j-1}^2$.
	To initiate a turbulent state in the system, we introduce white-in-time Gaussian pumping at site $p$ and a linear dissipation at sites $d$. For direct cascade, $p=10$, and for inverse cascade, we set $p=N-10$, where $N$ is the number of modes. The dissipated sites are always $1$,$2$,$N-1$ and $N$. The dynamical equations now read
	\[
	\Dot{a}_j=-i 2^{1+\frac{j}{2}} a_j^* a_{j+1} -i2^{\frac{j-1}{2}}a_{j-1}^2 - \sum_d\gamma_{j}a_j\delta_{jd} + \sqrt{P}\xi_j(t)\delta_{jp}\ ,
	\]
	where $\la\xi_p(t)\xi^*_p(t')\ra=\delta(t-t')$.
	The energy flux through the modes equals the rate of energy pumping into the system:
	\[
	\omega_pP=\Pi= \frac{\text{d}}{\text{d}t}\sum_{m=p}^j\omega_j|a_j|^2 = 4\omega_{j} V_j J_{j+1} ,
	\]
	where-
	$
	\begin{cases}
	\omega_j = 2^j\\
	V_j = 2^{\frac{j}{2}}
	\end{cases}
	$, and $J_j=\Im\{a_{j-1}^2a_j^*\}$. So we estimate $a_j\sim (\omega_pP)^{\frac{1}{3}}2^{-\frac{j+1}{2}}$.\\
	
	We transform to dimensionless fields $b_j=(\omega_pP)^{-\frac{1}{3}}2^{\frac{j+1}{2}}a_j$ and get the following equations for the $b_j$:
	\[
	(\omega_pP)^{-\frac{1}{3}}\Dot{b_j}=-i\lp b_j^* b_{j+1} + b_{j-1}^2\rp + \lp \omega_pP\rp^{-\frac{1}{6}}\sqrt{2}\xi(t)\delta_{jp}\ .
	\]
	After rescaling time by $\tau \equiv \lp \omega_pP\rp^\frac{1}{3}t$, the forcing term changes to $\la\xi_p(0) \xi^*_p(t)\ra=(\omega_pP)^{\frac{1}{3}}\delta(\tau)$. We define a dimensionless noise term $\Tilde{\xi_p}=\lp \omega_pP\rp^{-\frac{1}{6}}\xi_p$, such that $\la\Tilde{\xi}_i(0) \Tilde{\xi}_j^*(\tau)\ra = \sqrt{2}\delta_{ip}\delta_{jp}\delta(\tau)$. Now the equations take the form
	\be
	b_j'=-i\lp b_j^*b_{j+1}+b_{i-1}^2\rp- \sum_d\gamma b_j\delta_{jd}+\Tilde{\xi}_j\delta_{jp} \ .
	\label{eqn: doubling chain}
	\ee
	
	The only free parameter in equation \ref{eqn: doubling chain} is the value of dissipation. Similar to the Fibonacci chain, this value is empirically chosen to minimize the temperature build-up near the dissipation scale: for direct cascade $\gamma=1$ and in inverse cascade $\gamma=0.275$. The noise distribution is $\tilde{\xi}\sim\frac{1+i}{\sqrt{2\pi}}e^{-\frac{x^2}{2}}$. The time increment is $\text{d}t=0.01$, and the number of time steps is $4\times10^6$ for each of the $10^4$ iterations over which we average. The system is integrated by a $\text{4}^{\text{th}}$ order stochastic RK integrator, which has been written for this purpose.
	
	\subsection{Computation of cumulants}

	For every polynomial, defined above as
	$C_{kl\bar  s\ldots}(i)= b_{i-k}b_{i-l} b_{i-s}^*\ldots b_{j}^*$,
	the average $\langle C_{kl\bar  s\ldots}\rangle(j) $ is computed in post-processing. Angular brackets denote time-averaging. As mentioned in the main paper, due to the symmetry $b_j\mapsto-ib_j$ of the Hamiltonian parts in equations (\ref{eqn: fibo chain}-\ref{eqn: doubling chain}), for odd (even) orders only imaginary (real) parts of $\langle C_{kl\bar  s\ldots}\rangle(j) $ are non-zero.
	By double brackets we denote the cumulants, that
	is the mean values of the polynomials with all reducible contributions
	(if they exist) subtracted. For example, if two polynomials of orders
	$m_1$ and $m_2$ have nonzero mean values, from the mean value of the
	product polynomial $C_m = C_{m_1} C_{m_2}$ we need to subtract the
	product of the mean values multiplied by an appropriate combinatorial factor. As explained in the main paper, the ratio
	$\langle\langle C \rangle\rangle/\langle \tilde{C} \rangle$ scales as
	$1/|j-d|$ with the distance to damping.  Thus, we introduce
	compensated normalized cumulants,
	\[
	E_m(j) =  |j - d| \frac{\langle C_m(j) \rangle  - c_{\rm comb}
		\langle C_{m_1}(j) \rangle  \langle C_{m_2}(j) \rangle - \langle \dots \rangle
	}{\tilde{C}_m (j)}.
	\]
	Here $ \langle \dots \rangle$ denotes all possible decompositions into irreducible moments and $c_{\rm comb}$
	is the combinatorial factor of each decomposition.  For instance, in Fibonacci model, the moment
	$\bar{3}2111\bar{0}\bar{0}$ breaks down into $\bar{3}11\bar{0}$ and $21\bar{0}$ with
	$c_{\rm comb} = 6$, since each of two $\bar{0}$ and each of three $1$ can go into $21\bar{0}$.
	For unique decompositions we use notations $E_{m_1 m_2}$, such as $E_{43}$; for
	moments with multiple decompositions, $m = m_1 + m_2 = m_3+m_4$, we use
	$E_{m_1 m_2.m_3 m_4}$, such as $E_{43.52}$, etc; notaion $E_7$ stands for irreducible cumulants.
	The quantity in denominator is a normalization factor, based on the Gaussian fit
	$A_m(j) \approx \left( 1.13\, \Gamma(\frac{m}{2}+1)|j-d|\right)^{m/3}$, discussed in the main
	paper,
	\[
	\tilde{C}_{kl\bar  s\ldots}(j) = 1.13^{m/3} \sqrt{\mu}
	\left( |j-k-d| |j-l-d| |j-s-d| \ldots |j-d| \right )^{1/3},
	\]
	where $\mu$ is the product of factorials
	of numbers of appearances of each multiplier in the product, and $\la n_p\ra=1.13$ is the value of occupation number at the pumping scale.
	This normalization can also be expressed through
	time-averaged occupation numbers, $n_j = \langle|b_j|^2\rangle $, as
	$
	\tilde{C}_{kl\bar  s\ldots}(j)
	\approx \langle | b_{j-k}b_{j-k} b_{j-s}^*\ldots b_{j}^* |^2 \rangle^{1/2}
	\approx \left(\mu \, n_{j-k}n_{j-l} n_{j-s}\ldots n_{j}\right)^{1/2}.
	$
	We have confirmed that three methods of normalization give
	similar numerical results.
	
	Finding all the cumulants for an arbitrary $m$ is a nontrivial
	mathematical problem. In Fibonacci model, for $m=2$ there is only one
	nonzero moment $0\bar{0}$, the occupation numbers.
	The only triple moment, $21\bar{0}$, is responsible for the flux.
	For $m=4$ there are three irreducible moments, $332\bar{0}$,
	$431\bar{0}$, and $\bar{3}11\bar{0}$. For $m=5$, $6$, and $7$ we have counted
	8, 24, and 70 irreducible moments respectively.  Starting from $m=4$, irreducible
	moments are accompanied by an infinite number of reducible ones, for example
	$7\bar{7}21\bar{0}$ is formed by  $21\bar{0}$ and $7\bar{7}$,
	which is shifted $0\bar{0}$; similarly $1\bar{1}0\bar{0}$ is formed by
	$0\bar{0}$ and shifted $0\bar{0}$.  Note that since modes are counted down
	from any $i$, and to avoid redundancy by conjugation, the last digit is always
	$\bar{0}$. The far-shifted cumulants, however, tend to zero,
	as their components interact stronger when they overlap.
	%However, cumulants are nonzero
	%(at the main order in $|i-d|$ only when reducible moments have some
	%overlap in mode numbers, that is the same $b_j$ is encountered in both
	%reducible contributions.

	Analytic  consideration in the main text predict that  $E_m(j)$
	are constant in the inertial range of $j$ for the doubling model. Let us give respective arguments for the Fibonacci model. Nonzero correlation functions appear as the mean value of the polynomials 	$C_{kl\bar  s\ldots}(j)= b_{j-k}b_{j-l} b_{j-s}^*\ldots b_{j}^*$, which are
	gauge invariant: $F_{j-k}+F_{j-l}-F_{j-s}+\ldots=F_j$. Overbar in subscript  indicates complex conjugation for
	a mode in the product.\\
	
	Stationarity of the flux, $C_{21\bar0}$, gives: 
	\bea \!\!\!&i{ d\langle
		C_{21}(j) \rangle\over dt}=\phi^{-1}\langle
	|b_{j-1}b_j|^2\rangle+\langle |b_{j-2}b_j|^2\rangle-\phi\langle
	|b_{j-2}b_{j-1}|^2\rangle\nonumber\\ &+\langle C_{431\bar0}(j) \rangle
	\phi-\langle C_{431\bar0}(j+2) \rangle \phi^{-1}+\langle
	C_{322\bar0}(j) \rangle \phi\nonumber\\ &+\langle C_{322\bar0}(j+1)
	\rangle\left(\phi^{-1}-1\right)+C_{\bar311\bar0}\ .
	\label{Hopf1}
	\eea
	Using $\phi^2=1+\phi$, we approximate the first
	line:
	$$ n_{j-2}\bigl(n_j-n_{j-1}\bigr)+ {n_{j-1}\over\phi}\bigl(n_j-n_{j-2}\bigr)
	\approx\left(1+{2\over\phi}\right) {2\xi n_j^2\over|j-d|}\,,$$
	which is an estimate for the combination of cumulants of the fourth order,
	$\langle\langle C_4\rangle\rangle$. Computing time derivatives of higher
	$\langle C_m\rangle$, we come self-consistently to the same estimate
	for the dimensionless cumulants $D_m\propto |j-d|^{-1}$ and $E_m$ independent of $j$.\\

	This is supported by numerics, even though
	we see mode-to-mode fluctuations, especially for higher order moments, due to
	limited statistics. With this said, averaging over $i$ in the inertial range
	(we use $20 < i < 180$) provides reasonable metrics for  $E_m$.
	These values are shown in the tables below and used in the final ``sum of squares''
	$\sum{|E_m|^2}$.
	
	\newpage
	
	%-------------------------------------------
	\section{Details on the evaluation of cumulants}
	%-------------------------------------------
	
	\subsection{Estimations for the statistics - Numerical results}
	In this section we define $n_j = |a_j|^2$ and $J_j=b_{j-1}^2b_j^*$, i.e., without averages.
	\subsubsection{Logarithmic dependence of temperature on distance from dissipation}
	
\begin{figure}[h!]
		\centering
		\includegraphics[width=0.4\textwidth]{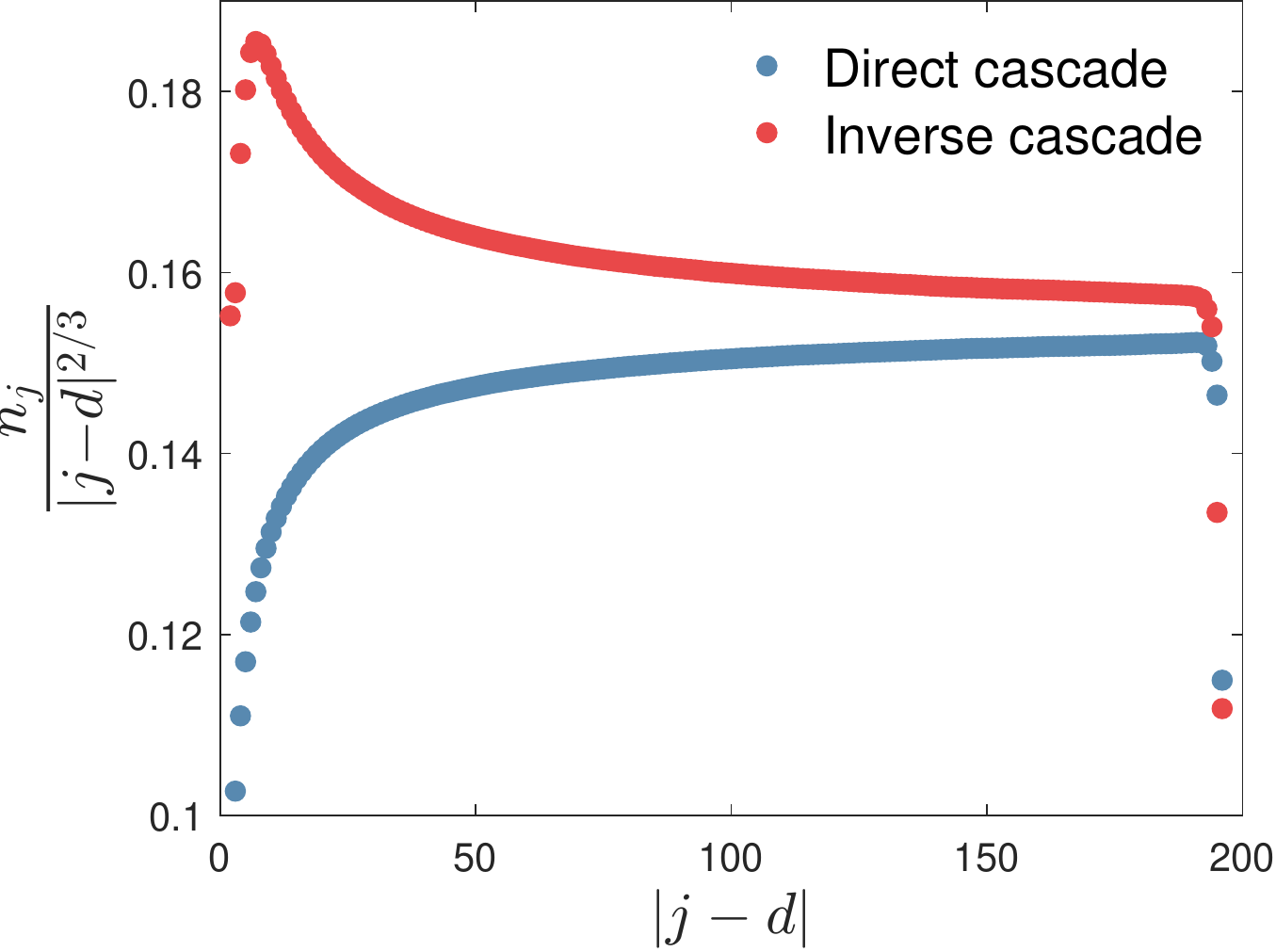}
		\includegraphics[width=0.4\textwidth]{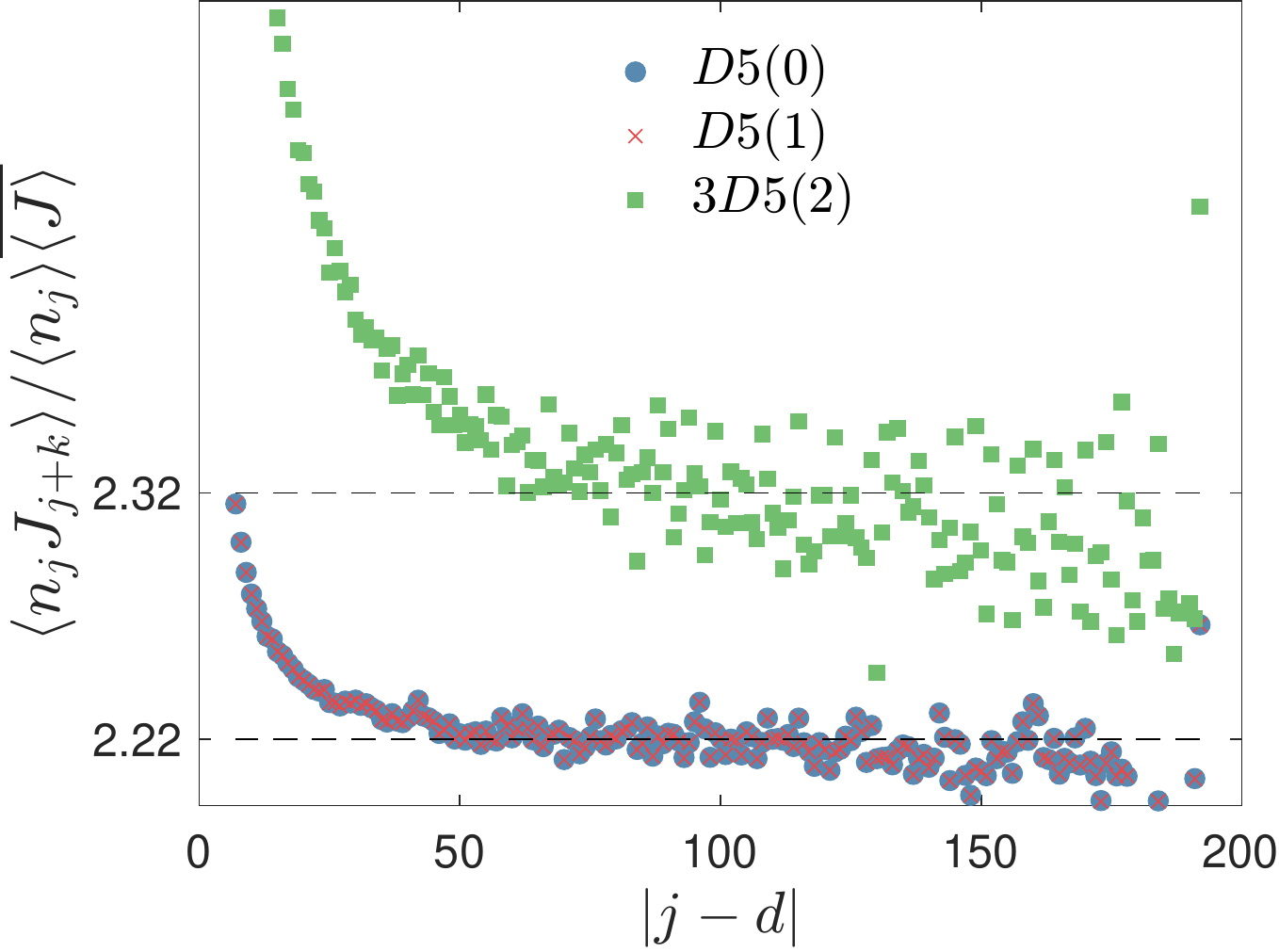}
		\caption{Left: Temperature dependence on distance from dissipation in a both cascades in doubling model.
			Right: Correlation function D5(k)=$\langle  n_j J_{j+k}\rangle$, normalized by $\la n_j\ra\overline{\la J\ra}$,
			where bar stands for averaging over the inertial range, for doubling model.
		}
		\label{Temperature}
		\label{normalized nJs}
	\end{figure}

	As discussed in the main text, the spectrum shows a logarithmic
	dependence on the distance from the damping scale, with scaling
	$\xi=\frac{2}{3}$, i.e. $\la n_j\ra\propto|j-d|^{\frac{2}{3}}$. This
	result is verified numerically and shown in the left panel of
	Fig.~\ref{Temperature} for both cascades in Doubling chain.
	Corresponding data for Fibonacci chain are included in main document.
	
	\subsubsection{Irreducibility of energy density and flux correlations in doubling cascade}
	
	Consider $D5(k)=\langle n_j J_{j+k}\rangle$. If it was reducible for all $k$, one expects $D5(1)=3D5(0)/2=3\la n_j\ra\la J_{j+1}\ra$.  Meanwhile, from the equation of motion for the fourth-order correlation function it is straightforward to derive $D5(0)=D5(1)=3D5(2)$. The right panel in Fig.~\ref{normalized nJs} shows the true values of the correlation functions:  $D5(0)=D5(1)=3D5(2)\approx 2.2\la n_j\ra\la J_{j+k}\ra$.
	
	From stationarity of the $\la n_j^{m+1}\ra$ moments, we get the identity:
	\be
	\la n_j^{m}J_{j+1}\ra = \la n_j^{m}J_j\ra \ .
	\label{eqn:nmJ}
	\ee
	This identity is clearly non-trivial as already seen above --- for $m=1$ it defies the basic expectation; for the polynomial PDF seen in the text in equations (10)-(12) we would estimate $\frac{\la n_j^{m}J_{j+1}\ra}{\la n_j^{m}J_{j}\ra}\sim\frac{\binom{m+2}{2}\la n_j\ra^{m}\overline{\la J\ra}}{\binom{m+1}{1}\la n_j\ra^{m}\overline{\la J\ra}}=\frac{m+2}{2}$ which is greater than 1 for all $m>0$. In general, it puts a strong constraint on the PDF, from which the consequences are yet to be fully extracted.
	
\begin{figure}[h!]
		\centering
		\includegraphics[width=0.4\textwidth]{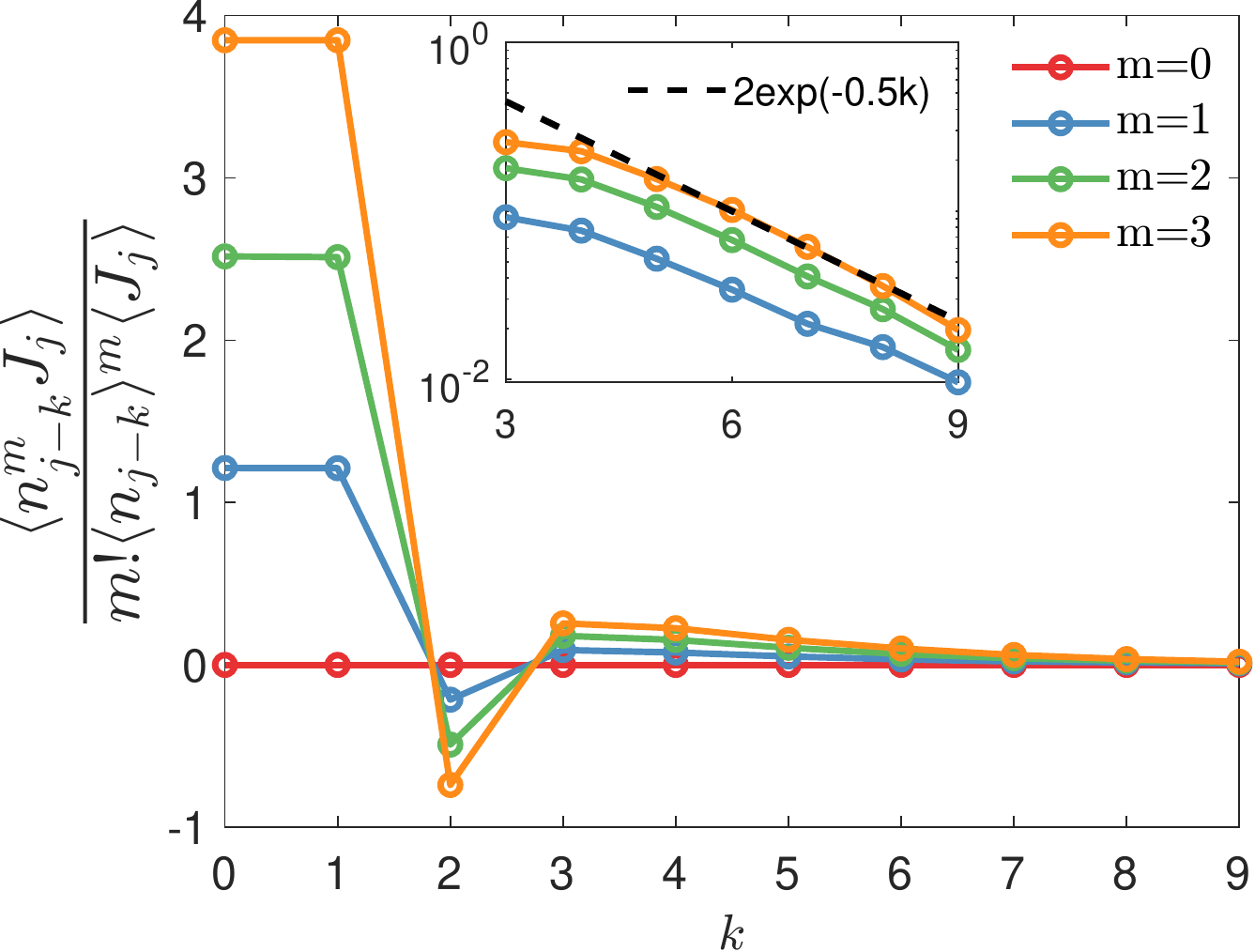}
		\includegraphics[width=0.4\textwidth]{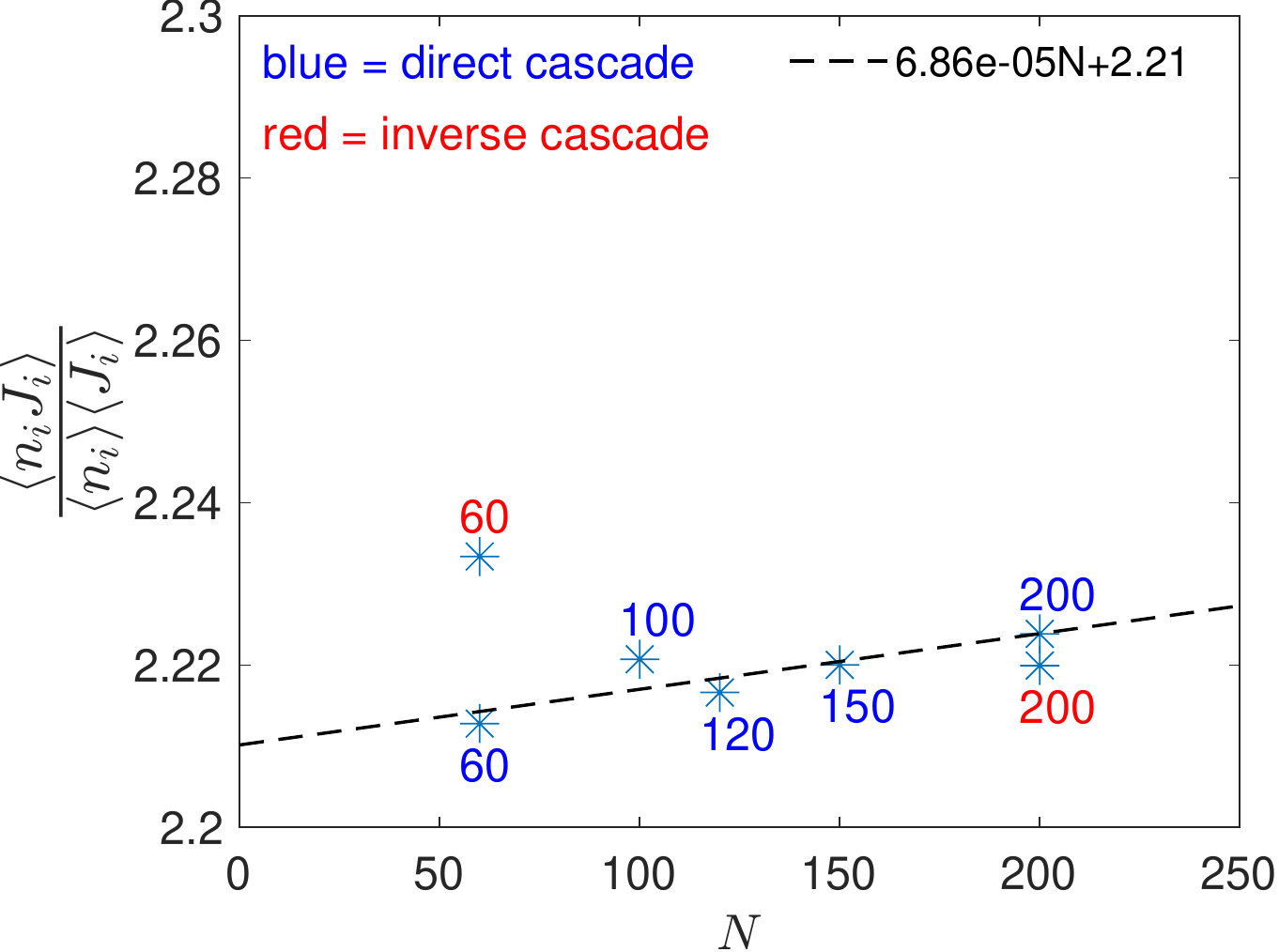}
		\caption{
			Left:
			Correlation between powers of energy density and the flux for $N=60$ in direct cascade. The irreducible parts tend exponentially fast to zero as the shift in indices grows.
			Right:
			Occupation number-flux correlations at the same site,
			normalized by multiplication of the means. These values are averaged over the inertial interval. The dashed line depicts a very moderate linear trend in the direct cascade data.
		}
		\label{dependence of nJ on distance}
		\label{nJ dependence on N}
	\end{figure}

	In the left panel of Fig.~\ref{dependence of nJ on distance} we validate identity (\ref{eqn:nmJ}) for $m=0,1,2,3$ in doubling system. In this figure, we also show that the irreducible parts of the $\la n_{j-k}^{m}J_{j}\ra$ cumulants tend to zero when there is no overlap. This decay is not monotonic with $k$ but has a small minimum for $k=2$, unlike the the decay of $\la J_{j-k}J_j\ra$ seen below. However, similar to the decay of $\la J_{j-k}J_j\ra$, the tails decay exponentially with $k$, and with a similar decay rate.
	It could be that on top of $|j-d|^{-1}$ dependence, there is an even weaker dependence on the interval width, see
	Fig.~\ref{nJ dependence on N}, right panel.

	\newpage
	
	\subsubsection{Flux fluctuations in doubling model}
	
\begin{figure}[h!]
		\centering
		\includegraphics[width=0.4\textwidth]{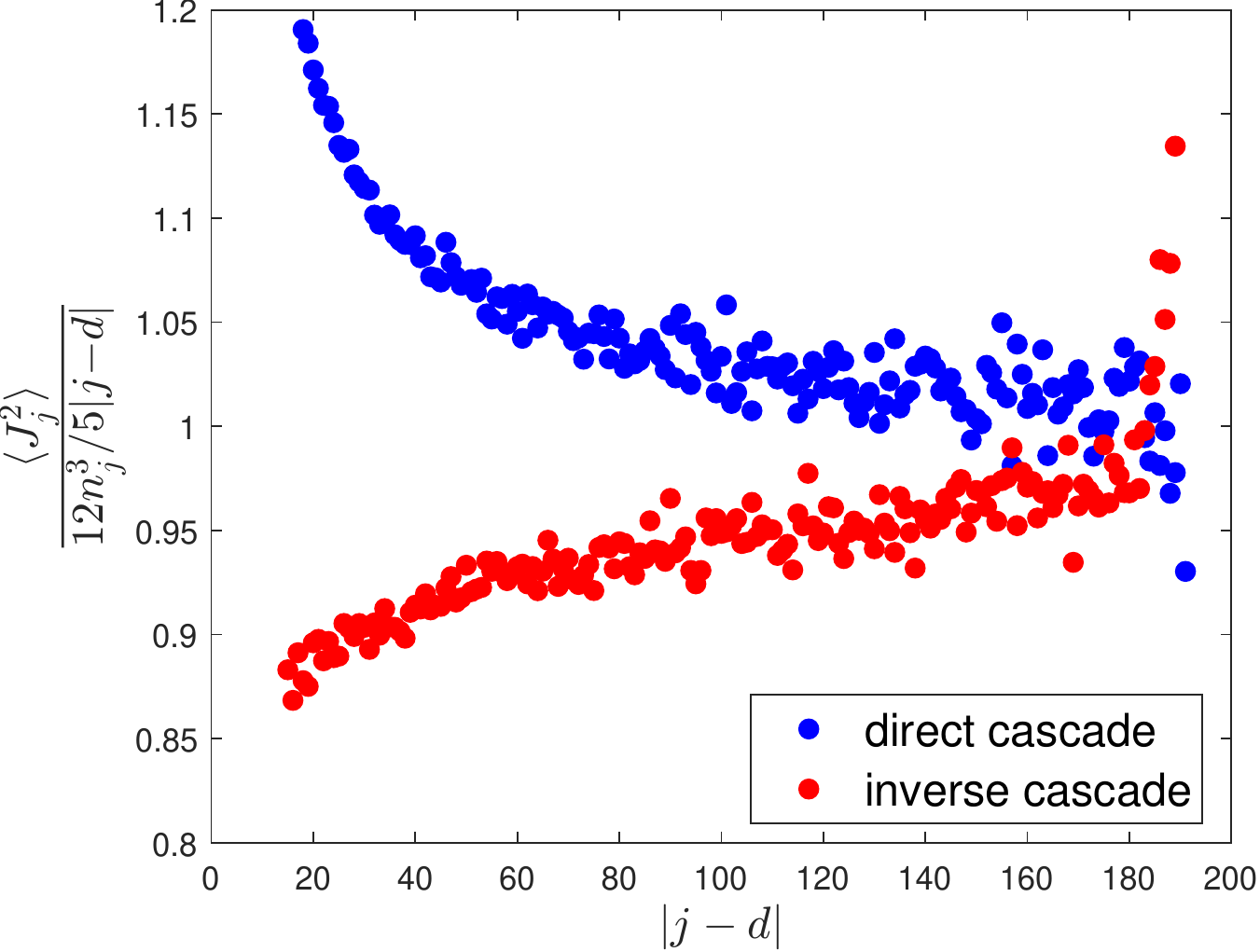}
		\includegraphics[width=0.4\textwidth]{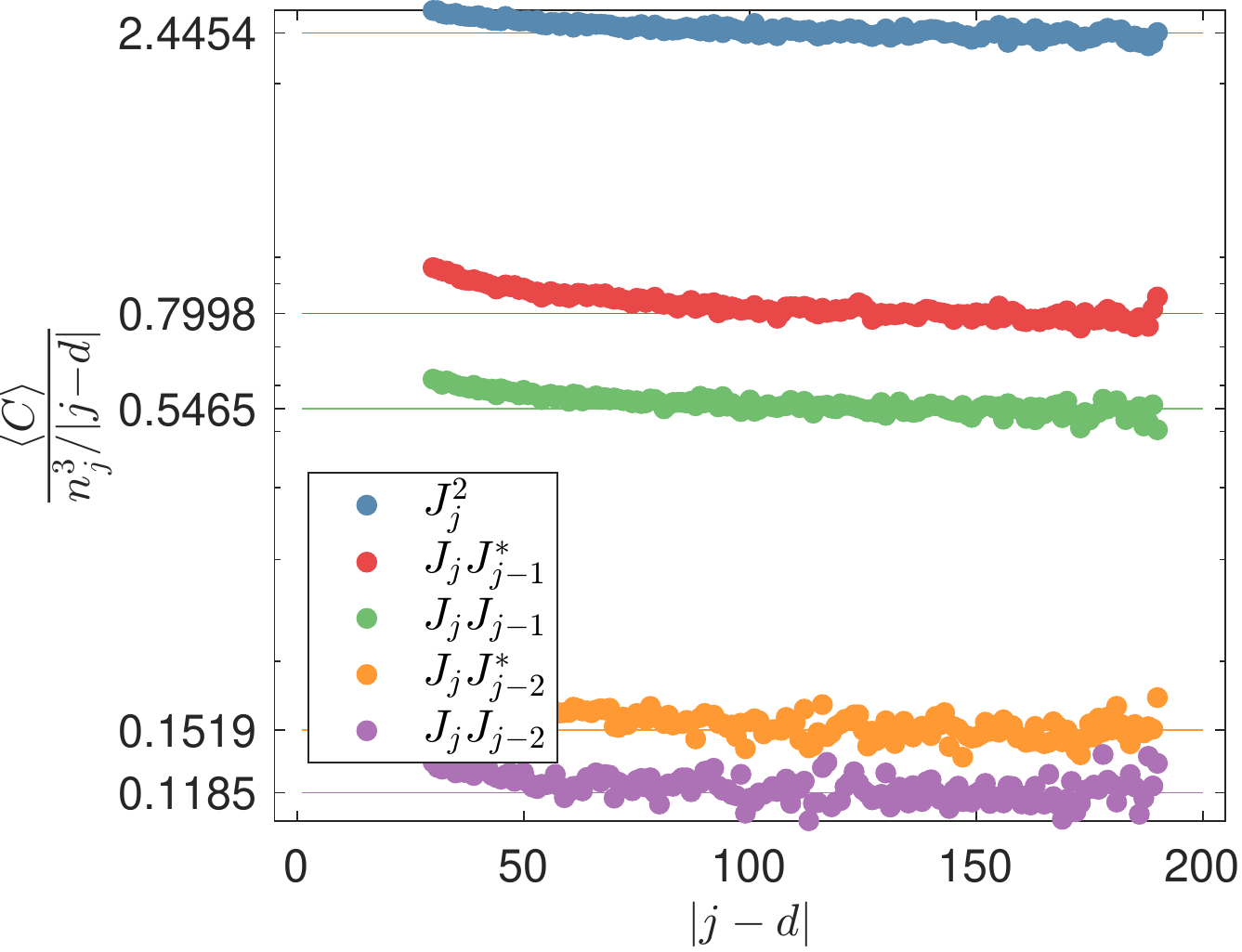}
		\caption{Left: The estimation $\langle J^2_j\rangle\approx\frac{2.4n_j^3}{|j-d|}$.
			Right: Values for the flux-flux correlators in direct cascade of doubling system.
		}
		\label{fig:flux variance}
		\label{fig:JJ values}
	\end{figure}

	The energy flux in turbulence is a stationary yet highly fluctuating quantity. From the dynamical equations on the fifth order cumulants, we get an estimation for the flux's variance, $\sigma_\Pi^2=\la J_j^2\ra\propto\frac{n_j^3}{|j-d|}\approx|j-d|$.  As shown in Fig.~\ref{fig:flux variance}, this estimation holds for both direct and inverse cascades, with the same proportionality constant $\approx2.4 $. In the normalized variables, $\overline{\Pi}=\frac{p}{N}$($\overline{\Pi}=\frac{N-p}{N}$) in direct(inverse) cascade, therefore $\sigma_\Pi \gg \overline{\Pi}$. This result and other flux-flux correlations, which are seen in Fig.~\ref{fig:JJ values}, show that the flux fluctuations are much larger than flux mean.

	\newpage
	
	\subsubsection{Decay of flux-flux fluctuations}
	
\begin{figure}[h!]
		\centering
		\includegraphics[width=7.5cm,height=6cm]{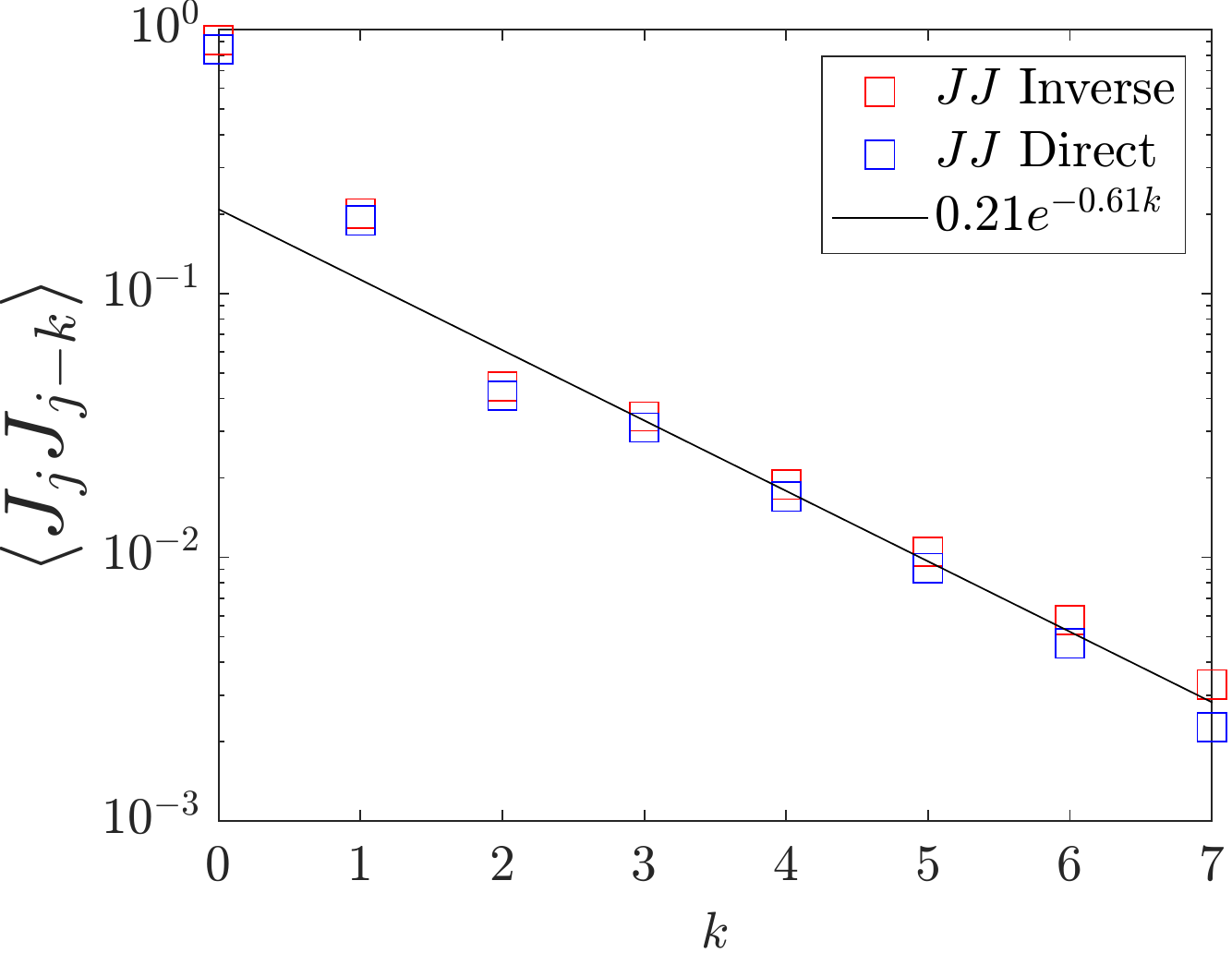}
		\includegraphics[width=7.5cm,height=6cm]{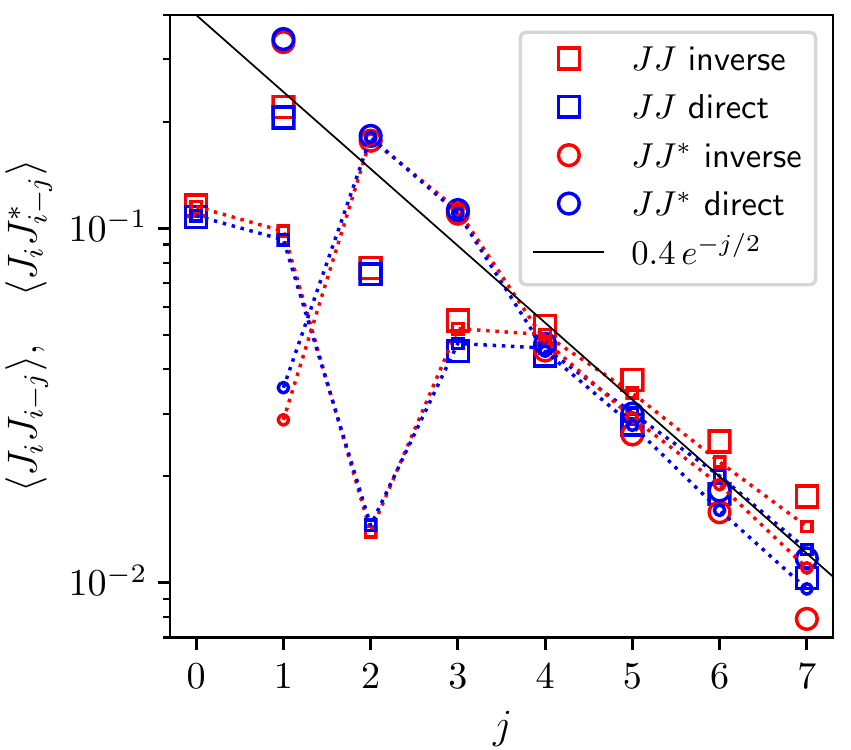}
		\caption{
			Left: Decay of flux-flux correlation in doubling system.
			Right: Fibonacci model.
			Large isolated symbols are showing the value of moments without adjustment for reducible part.
			On the left panel, small symbols connected with a dashed line show the values of adjusted moments;
			the drops at shifts $k=1$ and $k=2$ are due to subtraction of contributions of double-quadruple pairs.
			%The adjustment due to contributions of triple-triple pairs is insignificant.
		}
		\label{JJk}
		\label{fig:decay of flux-flux correlation}
	\end{figure}

	In Figure \ref{fig:decay of flux-flux correlation} we display the rate of decay of site-shifted flux-flux correlations for doubling and Fibonacci models.  As we  expected, the energy fluxes become exponentially less correlated with increasing shift in indices. In both models, we observe similar but not identical coefficients of exponential decay.

	Figure~\ref{JJk} shows that the flux is a strongly fluctuating
	quantity, that is
	$\langle ( b_{j-2} b_{j-1}b^*_j )^2\rangle\gg \langle (b_{j-2} b_{j-1}b^*_j )\rangle^2 $
	for the Fibinacci model and
	$\langle (b_{j-1}^2 b^*_j )^2\rangle\gg \langle (b_{j-1}^2 b^*_j )\rangle^2$
	for the doubling model.
	In particular, the main
	contribution into the sum of sixth-order cumulants is given by squared
	flux terms. Also for the doubling model, $\langle JJ\rangle$ terms
	give 0.1269, while all the other sixth-order cumulants give
	0.0733. Single-time correlations between flux values at different
	points beyond overlap decrease exponentially with the shift $k$ as
	$e^{-k/2}$, except $JJ(1)$.
	For Fibonacci model,
	$JJ(1) = \langle C_{21\bar0}(j) C_{21\bar0}(j+1)\rangle = (1+\phi)\langle C_{21\bar0}^2(j) \rangle/2$,
	which we derived analytically (from stationarity of $\langle n_j(J_j-J_{j+1})\rangle$) and confirmed numerically.

	\subsection{Relative entropy --- sum of squares in each order}
	The main difference between entropies of turbulence and thermal
	equilibrium is simply due to the distortion of equipartition, which
	lowers the entropy at a given total energy $\sum_jn_j$: the entropy of
	the Gaussian ${\cal P}_G(n_j)$ having distribution ${\cal P}_G(n_j)$
	with $n_j=5(j/N)^{2/3}$ is lower by $N[2/3-\ln (5/3)]$ than the entropy
	of thermal equilibrium with $n_j=1$. More informative is the
	difference between the turbulent distribution and the Gaussian
	distribution with the same occupation numbers. That difference is
	expressed in irreducible correlation functions called cumulants, which
	are zero for a Gaussian distribution.

	As explained in the main text, the sum of squares $\sum E_m^2$ measures the relative entropy between the Gaussian PDF ($\ln\mathcal{P}_G\propto-E$) and the polynomial PDF (equations (10)-(12) in the main text). In Tables \ref{tab:doubling relative entropy} and \ref{tab:fibonacci relative entropy} we count the multiplicity of such cumulants for each order and show their contribution to the sum of squares. That is the main result of our work.
	
	\begin{table*}
		\begin{tabular}{ |c|lr|cc|cc|cc|}
			\hline
			\makebox[15mm]{}   &  \makebox[15mm]{}  &  \makebox[10mm]{count}  & \makebox[15mm]{inverse} & \makebox[15mm]{direct} \\
			\hline
			$m=3$  &  $E_3$       &  1  &  0.05635   &  0.06343 \\
			\hline
			$m=4$  &  $E_4$       &  1  &  0.02540   &  0.02743 \\
			&  $E_{22}$     &  3   &  0.07599   &   0.04292 \\
			&  total       &  4   &  0.10140   &  0.07035 \\
			\hline
			$m=5$  &  $E_5$       &  2  &  0.00020   &  0.00026 \\
			&  $E_{32}$     &  4   &  0.00518   &   0.00568 \\
			&  total       &  6   &  0.00539   &  0.00594 \\
			\hline
			$m=6$  &  $E_6$        &  3  &  0.00449   &  0.00466 \\
			&  $E_{222}$     &  2   &  0.14040  &   0.06946 \\
			&  $E_{42}$     &  4   &  0.00060  &   0.00063 \\
			&  $E_{33}$     &  10  &  0.16785  &   0.18357 \\
			&  $E_{42.33}$  &   1   & 0.00034   &  0.00033  \\
			&  $E_{222.33}$  &   1   & 0.02502   &  0.03109  \\
			&  total       &   21  & 0.33869   &  0.28974  \\
			\hline
			$m=7$  & $E_7$         &  5   &    0.00112   &  0.00115  \\
			& $E_{322}$      &   9  &    0.01032   &  0.01118  \\
			& $E_{52}$      &   7  &    0.00023   &  0.00027  \\
			& $E_{43}$      &   5  &    0.00861   &  0.00946  \\
			& $E_{43.52}$    &  1  &    0.0001   &   0.00001 \\
			& $E_{43.322}$   &   2 &    0.00302    &  0.00315  \\
			&  total       &   29 &  0.02332      &  0.02521 \\
			\hline
			$m=8$  &  $E_8$       &  9  &  0.00106   &  0.00089 \\
			\hline
		\end{tabular}
		\caption{Partial sums of squares of cumulants computed in numerical simulations for the Doubling chain.
			For reducible moments count is the number of moments computed for partial sums.}
		\label{tab:doubling relative entropy}
	\end{table*}

	\begin{table*}
		\begin{tabular}{ |c|lr|cc|cc|cc|}
			\hline
			\makebox[15mm]{}   &  \makebox[15mm]{}  &  \makebox[10mm]{count}  & \makebox[15mm]{inverse} & \makebox[15mm]{direct} \\
			\hline
			$m=3$  &  $E_3$       &  1  &  0.189922   &  0.179861 \\
			\hline
			$m=4$  &  $E_4$       &  3  &  0.227484   &  0.226358 \\
			\hline
			$m=5$  &  $E_5$       &  8  &  0.048491  &   0.049622\\
			&  $E_{32}$               &  9   &  0.022531  &   0.021377 \\
			&  total                       &  17  &  0.071022  &  0.070999 \\
			\hline
			$m=6$  &  $E_6$        &  24  &  0.021736   &  0.021942 \\
			&  $E_{42}$                &  15 &   0.018720  &   0.017911 \\
			&  $E_{33}$                &  12  &  0.069856  &   0.067788 \\
			&  $E_{42.33}$           &   3   &  0.010556 &    0.010644 \\
			&  total                       &   54  & 0.120878  &  0.118285  \\
			\hline
			$m=7$  & $E_7$         &  70   &    0.008809  &   0.009171  \\
			& $E_{43}$                 &   56  &    0.027466   &   0.027042  \\
			& $E_{52}$                 &   40  &    0.008649   &   0.008244 \\
			& $E_{43.52}$            &  14  &     0.004217   &   0.004037 \\
			& $E_{43.43}$            &   1   &     0.000010   &   0.000019 \\			
			& $E_{43.43.52}$        &  2  &     0.002072   &   0.002025\\			
			& $E_{43.322}$           &   6 &    0.015772    &  0.014467 \\
			&  total                         &   189 &  0.066995      &  0.065005 \\
			\hline
		\end{tabular}
		\caption{Partial sums of squares of cumulants computed in numerical simulations for the Fibonacci chain.
			For reducible moments count show the number of moments computed for partial sums.}
		\label{tab:fibonacci relative entropy}
	\end{table*}

	In Table \ref{tab:doubling normalization table} we show the irreducible-normalized-squared form of all the cumulants in the doubling chain, i.e. $\frac{\lp\la C\ra-C_{irr}\rp^2}{C_{ms}}$, where $C_{irr}$ is the sum of irreducible contributions to $C$ and $C_{ms}$ is the mean-squared value of the cumulant for the Gaussian PDF. In the last column of the table, the mean value of the cumulant is to be substituted instead of the letter ``C''. The cumulants are specified uniquely by a symbol of the form $(kl...\bar{s})$ which corresponds to the number $b_{j-k}b_{j-l}...b_{j-s}^*$. $J$ is always defined by the complex number $J_j\equiv b_{j-1}^2b_j^*$, and a complex number of the form $b_{j-k}b_{j-l}...b_{j-s}^*b_{j}^*$ such that $2^{-k}+2^{-l}+...2^{-s}=1$, is denoted by $Ckl...\bar s$ (these are the irreducible cumulants).

	\begin{table}[h!]
		\small
		\caption {Table of irreducible-squared-normalized cumulants} \label{tab:doubling normalization table}
		\begin{center}
			\begin{tabular}{|cccc|}
				\hline
				order & symbol & definition & normalization\\
				\hline\hline
				3  &$(11\bar{0})$ & $\Im \langle b_{j-1}^2 b_j^* \rangle  $ & $C^2/2 n_{j-1}^2 n_j$ \\
				\hline\hline
				4  &$(221\bar{0})$ & $\Re \langle b_{j-2}^2 b_{j-1} b_j^* \rangle  $ &  $C^2/2n_{j-2}^2 n_{j-1} n_j$\\\cline{2-4}
				&$(00\bar{0}\bar{0})$ & $ \langle n_j^2\rangle  $ &  $(C-2n_j^2)^2/24n_{j}^4$\\\cline{2-4}
				&$(1\bar{1}0\bar{0})$ & $ \langle n_{j-1}n_j \rangle  $ &  $(C-n_{j-1}n_j)^2/4n^2_{j-1} n^2_j$\\\cline{2-4}
				&$(2\bar{2}0\bar{0})$ & $\langle n_{j-2}n_j \rangle  $ &  $(C-n_{j-2}n_j)^2/4n_{j-2}^2n^2_j$\\\cline{2-4}
				\hline\hline
				5  & $(3321\bar{0})$ & $\Im \langle b_{j-3}^2 b_{j-2} b_{j-1} b_j^* \rangle  $ & $C^2/2n_{j-3}^2 n_{j-2} n_{j-1} n_j$\\\cline{2-4}
				&$(2222\bar{0})$ & $\Im{\la b_{j-2}^4b_j^*\ra}$ & $C^2/24n_{j-2}^4 n_j$\\\cline{2-4}
				&$(110\bar{0}\bar{0})$ & $\Im{\la n_jJ_j\ra}$ & $(C-2n_jJ_j)^2/12n_{j-1}^2n_j^3$\\\cline{2-4}
				&$(111\bar{1}\bar{0})$ & $\Im{\la n_{j-1}J_j\ra}$ & $(C-3n_{j-1}J_j)^2/12n_{j-1}^2n_j^3$\\\cline{2-4}
				\hline\hline
				6  &$(44321\bar{0})$ & $\Re \langle b_{j-4}^2 b_{j-3} b_{j-2} b_{j-1} b_j^* \rangle  $ & $C^2/2n_{j-4}^2 n_{j-3} n_{j-2} n_{j-1} n_j$\\\cline{2-4}
				
				&$(33331\bar{0})$ & $\Re \langle b_{j-3}^4 b_{j-1} b_j^* \rangle  $ & $C^2/24n_{j-3}^4 n_{j-1} n_j$\\\cline{2-4}
				
				&$(33222\bar{0})$ & $\Re \langle b_{j-3}^2 b_{j-2}^3 b_j^* \rangle  $ & $C^2/12n_{j-3}^2 n_{j-2}^3 n_j$\\\cline{2-4}
				
				&$(1111\bar{0}\bar{0})$ & $\Re{\la J_{j}^2\ra}$ & $(C-6J_j^2)^2/48n_{j-1}^4n_j^2$\\\cline{2-4}
				
				&$(11\bar{1}\bar{1}0\bar{0})$ & $\Re\la J_jJ_j^*\ra$ & $(C-2n_{j-1}^2n_j-J_j^2)^2/48n_{j-1}^4n_j^2$\\\cline{2-4}
				
				&$(2211\bar{1}\bar{0})$ & $\Re{\la J_{j}J_{j-1}\ra}$ & $(C-2C221_jn_{j-1}-2J_j^2)^2/12n_{j-2}^2n_{j-1}^3n_j$\\\cline{2-4}
				
				&$(\bar{2}\bar{2}111\bar{0})$ & $\Re{\la J_{j}J^*_{j-1}\ra}$ & $(C-3J_j^2)^2/12n_{j-2}^2n_{j-1}^3n_j$\\\cline{2-4}
				
				&$(33\bar{2}11\bar{0})$ & $\Re{\la J_{j}J_{j-2}\ra}$ & $(C-J_j^2)^2/4n_{j-3}^2n_{j-2}n_{j-1}^2n_j$\\\cline{2-4}
				
				&$(\bar{3}\bar{3}211\bar{0})$ & $\Re{\la J_{j}J_{j-2}\ra}$ & $(C-J_j^2)^2/4n_{j-3}^2n_{j-2}n_{j-1}^2n_j$\\\cline{2-4}
				
				&$(2210\bar{0}\bar{0})$ & $\Re{\la C221_jn_j\ra}$ & $(C-2C221_jn_j)^2/12n_{j-2}^2n_{j-1}n_j^3$\\\cline{2-4}
				
				&$(222\bar{2}1\bar{0})$ & $\Re{\la C221_jn_{j-2}\ra}$ & $(C-3C221_jn_{j-2})^2/24n_{j-2}^4n_{j-1}n_j$\\\cline{2-4}
				\hline\hline
				7 &$(554321\bar{0})$ & $\Im \langle b_{j-5}^2 b_{j-4} b_{j-3} b_{j-2} b_{j-1} b_j^* \rangle  $ & $C^2/2n_{j-5}^2 n_{j-4} n_{j-3} n_{j-2} n_{j-1} n_j$\\\cline{2-4}
				
				&$(444421\bar{0})$ & $\Im \langle b_{j-4}^4 b_{j-2} b_{j-1} b_j^* \rangle  $ & $C^2/24n_{j-4}^4 n_{j-2} n_{j-1} n_j$\\\cline{2-4}
				
				&$(443331\bar{0})$ & $\Im \langle b_{j-4}^2 b_{j-3}^3 b_{j-1} b_j^* \rangle  $ & $C^2/12n_{j-4}^2 n_{j-3}^3 n_{j-1} n_j$\\\cline{2-4}
				
				&$(443222\bar{0})$ & $\Im \langle b_{j-4}^2 b_{j-3} b_{j-2}^3 b_j^* \rangle  $ & $C^2/12n_{j-4}^2 n_{j-3} n_{j-2}^3 n_j$\\\cline{2-4}
				
				&$(333322\bar{0})$ & $\Im \langle b_{j-3}^4 b_{j-2}^2 b_j^* \rangle  $ & $C^2/48n_{j-3}^4 n_{j-2}^2 n_j$\\\cline{2-4}
				
				&$(1100\bar{0}\bar{0}\bar{0})$ & $\Im{\la n_j^2J_j\ra}$ & $(C-6n_j^2J_j)^2/240n_{j-1}^2n_{j}^5$\\\cline{2-4}
				
				&$(111\bar{1}0\bar{0}\bar{0})$ & $\Im{\la n_{j-1}n_jJ_j\ra}$ & $(C-6n_{j-1}n_jJ_ j)^2/144n_{j-1}^4n_{j}^3$\\\cline{2-4}
				
				&$(1111\bar{1}\bar{1}\bar{0})$ & $\Im{\la n_{j-1}^2J_j\ra}$ & $(C-12n_{j-1}^2J_j)^2/720n_{j-1}^6n_{j}$\\\cline{2-4}
				
				&$(2\bar{2}111\bar{1}\bar{0})$ & $\Im{\la n_{j-2}n_{j-1}J_j\ra}$ & $(C-3n_{j-2}n_{j-1}J_j)^2/48n_{j-2}^2n_{j-1}^4n_j$\\\cline{2-4}
				
				&$(22\bar{2}\bar{2}11\bar{0})$ & $\Im{\la n_{j-2}^2J_j\ra}$ & $(C-2n_{j-2}^2J_j+2C221J)^2/48n_{j-2}^4n_{j-1}^2n_j$\\\cline{2-4}
				
				&$(11\bar{1}0\bar{0}\bar{0})$ & $\Im{\la n_{j-1}n_jJ_{j-1}\ra}$ & $(C-2n_{j-1}n_jJ_j+2C221J)^2/24n_{j-2}^2n_{j-1}^3n_j^2$\\\cline{2-4}
				
				&$(33\bar{2}1\bar{1}0\bar{0})$ & $\Im{\la n_{j-1}n_jJ_{j-2}\ra}$ & $(C-n_{j-1}n_jJ_{j})^2/8n_{j-3}^2n_{j-2}n_{j-1}^2n_j^2$\\\cline{2-4}
				
				&$(332\bar{2}\bar{2}0\bar{0})$ & $\Im{\la n_{j-2}n_jJ_{j-2}\ra}$ & $(C-2n_{j-2}n_jJ_{j})^2/24n_{j-3}^2n_{j-2}^3n_j^2$\\\cline{2-4}
				
				&$(44\bar{3}2\bar{2}0\bar{0})$ & $\Im{\la n_{j-2}n_jJ_{j-3}\ra}$ & $(C-n_{j-2}n_jJ_{j})^2/8n_{j-4}^2n_{j-3}n_{j-2}^2n_{j}^2$\\\cline{2-4}
				
				&$(443\bar{3}\bar{3}0\bar{0})$ & $\Im{\la n_{j-3}n_jJ_{j-3}\ra}$ & $(C-2n_{j-3}n_jJ_{j})^2/24n_{j-4}^2n_{j-3}^3n_j^2$\\\cline{2-4}
				
				&$(22220\bar{0}\bar{0})$ & $\Im{\la n_jC2222_j\ra}$ & $(C-2n_jC2222_j)^2/144n_{j-2}^4n_j^3$\\\cline{2-4}
				
				&$(22221\bar{1}\bar{0})$ & $\Im{\la n_{j-1}C2222_j\ra}$ & $(C-n_{j-1}C2222_j-6C221J)^2/48n_{j-2}^4n_{j-1}^2n_j$\\\cline{2-4}
				
				&$(22222\bar{2}\bar{0})$ & $\Im{\la n_{j-2}C2222_j\ra}$ & $(C-5n_{j-2}C2222_j)^2/720n_{j-2}^6n_j$\\\cline{2-4}
				
				&$(3\bar{3}2222\bar{0})$ & $\Im{\la n_{j-3}C2222_j\ra}$ & $(C-n_{j-3}C2222_j)^2/48n_{j-3}^2n_{j-2}^4n_j$\\\cline{2-4}
				
				&$(33210\bar{0}\bar{0})$ & $\Im{\la n_jC3321_j\ra}$ & $(C-2n_jC3321_j)^2/12n_{j-3}^2n_{j-2}n_{j-1}n_j^3$\\\cline{2-4}
				
				&$(33211\bar{1}\bar{0})$ & $\Im{\la n_{j-1}C3321_j\ra}$ & $(C-2n_{j-1}C3321_j)^2/12n_{j-3}^2n_{j-2}n_{j-1}^3n_j$\\\cline{2-4}
				
				&$(3322\bar{2}1\bar{0})$ & $\Im{\la n_{j-2}C3321_j\ra}$ & $(C-2n_{j-2}C3321_j)^2/12n_{j-3}^2n_{j-2}^2n_{j-1}n_j$\\\cline{2-4}
				
				&$(333\bar{3}21\bar{0})$ & $\Im{\la n_{j-3}C3321_j\ra}$ & $(C-3n_{j-3}C3321_j)^2/24n_{j-3}^4n_{j-2}n_{j-1}n_j$\\\cline{2-4}
				
				&$(22111\bar{0}\bar{0})$ & $\Im{\la C221_jJ_{j}\ra}$ & $(C-6C221_jJ_{j})^2/24n_{j-2}^2n_{j-1}^3n_j^2$\\\cline{2-4}
				
				&$(3322\bar{2}1\bar{0})$ & $\Im{\la C221_jJ_{j-2}\ra}$ & $(C-C221_jJ_{j})^2/12n_{j-3}^2n_{j-2}^3n_{j-1}n_j$\\\cline{2-4}
				
				&$(\bar{3}\bar{3}2221\bar{0})$ & $\Im{\la C221_jJ^*_{j-2}\ra}$ & $(C+3C221_jJ_{j})^2/12n_{j-3}^2n_{j-2}^3n_{j-1}n_j$\\\cline{2-4}
				
				&$(44\bar{3}221\bar{0})$ & $\Im{\la C221_jJ_{j-3}\ra}$ & $(C-C221_jJ_{j})^2/4n_{j-4}^2n_{j-3}n_{j-2}^2n_{j-1}n_j$\\\cline{2-4}
				
				&$(\bar{4}\bar{4}3221\bar{0})$ & $\Im{\la C221_jJ^*_{j-3}\ra}$ & $(C+C221_jJ_{j})^2/4n_{j-4}^2n_{j-3}n_{j-2}^2n_{j-1}n_j$\\\cline{2-4}
				\hline\hline
				9  &$(111111\bar{0}\bar{0}\bar{0})$ & $\Im{\la J_{j}^3\ra}$ & $(C-90J_j^3)^2/4320n_{j-1}^6n_j^3$\\\cline{2-4}
				\hline\hline
				12  &$(11111111\bar{0}\bar{0}\bar{0}\bar{0})$ & $\Re{\la J_{j}^4\ra}$ & $(C-2520J_j^4)^2/4!8!n_{j-1}^8n_j^4$\\\cline{2-4}
				\hline\hline
			\end{tabular}
		\end{center}
	\end{table}

	%-------------------------------------

\section{Appendix: Tables of correlators for Fibonacci chain}

In the tables we use ``text'' notaions where the digits after dash stands for
conjugated modes, for example {\tt  611-440} is equivalent to $6\bar{4}\bar{4}11\bar{0}$.

\begin{footnotesize}
\begin{verbatim}
--------------------------------------------------------------------------------------------------------------------
IRREDUCIBLE MOMENTS      DIRECT                            INVERSE
--------------------------------------------------------------------------------------------------------------------
                     
-- IRREDUCIBLE  E3 --
 
21-0                -0.4241 +/- 0.0040                +0.4358 +/- 0.0045

SUM OF SQUARES E3 (1):    0.179861                          0.189922 

-- IRREDUCIBLE  E4 --

322-0               -0.1634 +/- 0.0048                +0.1707 +/- 0.0049
431-0               -0.0879 +/- 0.0040                +0.0869 +/- 0.0038
11-30               +0.4381 +/- 0.0047                -0.4368 +/- 0.0050

SUM OF SQUARES E4 (3):     0.226358                        0.227484

-- IRREDUCIBLE  E5 --

4332-0               -0.0712 +/- 0.0033                +0.0722 +/- 0.0027
5422-0               -0.0538 +/- 0.0036                +0.0589 +/- 0.0045
5441-0               -0.0171 +/- 0.0036                +0.0152 +/- 0.0033
6531-0               -0.0214 +/- 0.0036                +0.0224 +/- 0.0036

411-20               -0.1346 +/- 0.0040                +0.1279 +/- 0.0037
331-50               -0.0146 +/- 0.0035                +0.0144 +/- 0.0036
222-40               -0.0351 +/- 0.0032                +0.0354 +/- 0.0034
11-540               +0.1461 +/- 0.0031                -0.1458 +/- 0.0041

SUM OF SQUARES E5 (8):    0.0496218                         0.0484914

-- IRREDUCIBLE E6 --

44333-0               -0.0260 +/- 0.0041                +0.0264 +/- 0.0037
54432-0               -0.0404 +/- 0.0035                +0.0418 +/- 0.0039
65332-0               -0.0107 +/- 0.0036                +0.0089 +/- 0.0036
65522-0               -0.0225 +/- 0.0037                +0.0277 +/- 0.0045
76422-0               -0.0140 +/- 0.0039                +0.0148 +/- 0.0034

76631-0               -0.0101 +/- 0.0032                +0.0120 +/- 0.0036
63333-0               +0.0094 +/- 0.0037                -0.0098 +/- 0.0041
65541-0               -0.0085 +/- 0.0041                +0.0082 +/- 0.0037
76441-0               +0.0009 +/- 0.0036                -0.0017 +/- 0.0037
87531-0               -0.0039 +/- 0.0041                +0.0036 +/- 0.0038

3332-50               +0.0298 +/- 0.0037                -0.0265 +/- 0.0037
4422-60               +0.0223 +/- 0.0038                -0.0211 +/- 0.0036
4441-60               +0.0054 +/- 0.0041                -0.0075 +/- 0.0043
5531-70               +0.0042 +/- 0.0037                -0.0049 +/- 0.0035
6511-20               -0.0462 +/- 0.0039                +0.0438 +/- 0.0037

6331-40               +0.0245 +/- 0.0038                -0.0281 +/- 0.0040
5222-30               +0.0309 +/- 0.0045                -0.0398 +/- 0.0062
4111-00               +0.0629 +/- 0.0037                -0.0617 +/- 0.0034
611-440               -0.0503 +/- 0.0041                +0.0483 +/- 0.0036
311-520               +0.0468 +/- 0.0048                -0.0370 +/- 0.0047

331-760               -0.0188 +/- 0.0038                +0.0181 +/- 0.0038
222-650               -0.0385 +/- 0.0038                +0.0400 +/- 0.0038
11-7640               +0.0334 +/- 0.0036                -0.0329 +/- 0.0032
11-6550               +0.0423 +/- 0.0037                -0.0433 +/- 0.0038

SUM OF SQUARES E6 (24):       0.0219424                      0.0217464

-- IRREDUCIBLE E7 --

544433-0               -0.0215 +/- 0.0036                +0.0222 +/- 0.0040
554442-0               -0.0114 +/- 0.0033                +0.0116 +/- 0.0032
654333-0               -0.0005 +/- 0.0034                -0.0002 +/- 0.0035
655432-0               -0.0163 +/- 0.0036                +0.0178 +/- 0.0034
665551-0               -0.0023 +/- 0.0037                +0.0020 +/- 0.0039
744442-0               -0.0007 +/- 0.0034                -0.0001 +/- 0.0035
764432-0               -0.0112 +/- 0.0038                +0.0112 +/- 0.0035
766332-0               -0.0004 +/- 0.0039                -0.0007 +/- 0.0034
766522-0               -0.0141 +/- 0.0038                +0.0174 +/- 0.0036
766541-0               -0.0024 +/- 0.0036                +0.0024 +/- 0.0034
855551-0               -0.0014 +/- 0.0041                +0.0012 +/- 0.0035
873333-0               +0.0048 +/- 0.0032                -0.0046 +/- 0.0036
875332-0               -0.0018 +/- 0.0032                +0.0015 +/- 0.0039
875522-0               -0.0047 +/- 0.0035                +0.0058 +/- 0.0039
875541-0               -0.0016 +/- 0.0032                +0.0011 +/- 0.0033
877422-0               -0.0028 +/- 0.0034                +0.0023 +/- 0.0035
877441-0               +0.0022 +/- 0.0040                -0.0035 +/- 0.0035
877631-0               -0.0055 +/- 0.0039                +0.0069 +/- 0.0032
986422-0               -0.0032 +/- 0.0035                +0.0027 +/- 0.0036
986441-0               +0.0010 +/- 0.0035                -0.0010 +/- 0.0035
986631-0               -0.0030 +/- 0.0036                +0.0037 +/- 0.0039
988531-0               -0.0001 +/- 0.0039                -0.0005 +/- 0.0034
A97531-0               +0.0002 +/- 0.0181                +0.0008 +/- 0.0194
                       
43333-50               +0.0196 +/- 0.0037                -0.0170 +/- 0.0038
44432-60               +0.0178 +/- 0.0035                -0.0178 +/- 0.0040
52222-10               -0.0166 +/- 0.0035                +0.0158 +/- 0.0028
53333-70               -0.0017 +/- 0.0036                +0.0016 +/- 0.0038
55332-70               +0.0003 +/- 0.0043                -0.0002 +/- 0.0038
55522-70               +0.0081 +/- 0.0035                -0.0095 +/- 0.0040
55541-70               +0.0005 +/- 0.0034                +0.0002 +/- 0.0032
63331-20               -0.0139 +/- 0.0035                +0.0131 +/- 0.0036
63332-40               -0.0068 +/- 0.0032                +0.0052 +/- 0.0033
65111-00               +0.0256 +/- 0.0038                -0.0243 +/- 0.0036
66422-80               +0.0013 +/- 0.0038                -0.0008 +/- 0.0036
66441-80               -0.0024 +/- 0.0038                +0.0015 +/- 0.0040
66631-80               +0.0050 +/- 0.0040                -0.0059 +/- 0.0038
74422-50               -0.0076 +/- 0.0036                +0.0069 +/- 0.0035
74441-50               -0.0026 +/- 0.0038                +0.0035 +/- 0.0033
76222-30               +0.0162 +/- 0.0038                -0.0205 +/- 0.0041
76611-20               -0.0122 +/- 0.0037                +0.0122 +/- 0.0035
77531-90               -0.0004 +/- 0.0037                +0.0010 +/- 0.0036
87331-40               +0.0113 +/- 0.0036                -0.0130 +/- 0.0039
87511-20               -0.0111 +/- 0.0037                +0.0109 +/- 0.0031
85531-60               -0.0029 +/- 0.0041                +0.0034 +/- 0.0036

3111-500               -0.0220 +/- 0.0038                +0.0174 +/- 0.0036
3332-760               +0.0071 +/- 0.0035                -0.0070 +/- 0.0032
4222-630               -0.0142 +/- 0.0034                +0.0154 +/- 0.0040
4422-870               +0.0086 +/- 0.0036                -0.0079 +/- 0.0032
4441-870               +0.0027 +/- 0.0033                -0.0045 +/- 0.0036
5331-740               -0.0065 +/- 0.0034                +0.0065 +/- 0.0037
5511-720               +0.0155 +/- 0.0036                -0.0144 +/- 0.0039
5531-980               +0.0013 +/- 0.0041                -0.0024 +/- 0.0031
6311-420               -0.0160 +/- 0.0038                +0.0129 +/- 0.0035
7222-550               +0.0283 +/- 0.0037                -0.0281 +/- 0.0037
8331-660               +0.0130 +/- 0.0032                -0.0119 +/- 0.0035
8711-440               -0.0114 +/- 0.0035                +0.0113 +/- 0.0038
9621-770               -0.0085 +/- 0.0037                +0.0036 +/- 0.0045

111-5220               -0.0251 +/- 0.0038                +0.0232 +/- 0.0033
222-7660               -0.0149 +/- 0.0034                +0.0161 +/- 0.0035
222-8750               -0.0154 +/- 0.0029                +0.0162 +/- 0.0035
311-7620               +0.0194 +/- 0.0035                -0.0155 +/- 0.0036
331-8770               -0.0076 +/- 0.0036                +0.0077 +/- 0.0036
331-9860               -0.0073 +/- 0.0035                +0.0072 +/- 0.0038
511-7440               +0.0097 +/- 0.0044                -0.0060 +/- 0.0041
711-5550               -0.0172 +/- 0.0032                +0.0169 +/- 0.0034
811-6640               -0.0067 +/- 0.0038                +0.0069 +/- 0.0039

11-76650               +0.0216 +/- 0.0029                -0.0218 +/- 0.0037
11-87550               +0.0067 +/- 0.0039                -0.0075 +/- 0.0033
11-87740               +0.0084 +/- 0.0034                -0.0081 +/- 0.0036
11-98640               +0.0055 +/- 0.0036                -0.0056 +/- 0.0037

SUM OF SQUARES E7 (70):       0.0091714                      0.00880917 

--------------------------------------------------------------------------------------------------------------------
MOMENTS WITH SINGLE COMPOSITION                 DIRECT                            INVERSE
--------------------------------------------------------------------------------------------------------------------
-- REDUCIBLE  E32 --

210-00 (21-0)+(0-0) x2   (-0.3334) - (-0.3487) = +0.0154 +/- 0.0023    (+0.3399) - (+0.3562) = -0.0163 +/- 0.0021
211-10 (21-0)+(1-1) x2   (-0.3657) - (-0.3487) = -0.0170 +/- 0.0022    (+0.3748) - (+0.3562) = +0.0185 +/- 0.0029
221-20 (21-0)+(2-2) x2   (-0.2807) - (-0.3487) = +0.0681 +/- 0.0035    (+0.2838) - (+0.3563) = -0.0725 +/- 0.0019
321-30 (21-0)+(3-3) x1   (-0.2355) - (-0.3020) = +0.0665 +/- 0.0031    (+0.2384) - (+0.3086) = -0.0702 +/- 0.0024
421-40 (21-0)+(4-4) x1   (-0.3069) - (-0.3020) = -0.0049 +/- 0.0032    (+0.3104) - (+0.3086) = +0.0018 +/- 0.0021

320-10 (32-1)+(0-0) x1   (-0.3492) - (-0.2980) = -0.0512 +/- 0.0028    (+0.3651) - (+0.3130) = +0.0521 +/- 0.0031
430-20 (43-2)+(0-0) x1   (-0.2062) - (-0.2941) = +0.0879 +/- 0.0022    (+0.2328) - (+0.3177) = -0.0850 +/- 0.0028
540-30 (54-3)+(0-0) x1   (-0.3274) - (-0.2904) = -0.0370 +/- 0.0021    (+0.3650) - (+0.3227) = +0.0423 +/- 0.0047
650-40 (65-4)+(0-0) x1   (-0.2797) - (-0.2868) = +0.0071 +/- 0.0022    (+0.3253) - (+0.3279) = -0.0026 +/- 0.0044

SUM OF SQUARES E32 (9):                  0.0213773                                      0.0225309

-- REDUCIBLE E42 --       

3220-00 (322-0)+(0-0) x2   (-0.1090) - (-0.1344) = +0.0254 +/- 0.0024    (+0.1124) - (+0.1395) = -0.0271 +/- 0.0021
3222-20 (322-0)+(2-2) x3   (-0.1150) - (-0.1425) = +0.0275 +/- 0.0024    (+0.1173) - (+0.1480) = -0.0307 +/- 0.0020
3322-30 (322-0)+(3-3) x2   (-0.1050) - (-0.1344) = +0.0294 +/- 0.0026    (+0.1114) - (+0.1396) = -0.0282 +/- 0.0024
4322-40 (322-0)+(4-4) x1   (-0.1012) - (-0.1164) = +0.0152 +/- 0.0028    (+0.1051) - (+0.1209) = -0.0157 +/- 0.0027  
5322-50 (322-0)+(5-5) x1   (-0.1054) - (-0.1163) = +0.0110 +/- 0.0028    (+0.1071) - (+0.1209) = -0.0138 +/- 0.0027  

4310-00 (431-0)+(0-0) x2   (-0.0598) - (-0.0723) = +0.0125 +/- 0.0020    (+0.0590) - (+0.0710) = -0.0120 +/- 0.0021
4311-10 (431-0)+(1-1) x2   (-0.0502) - (-0.0723) = +0.0220 +/- 0.0020    (+0.0492) - (+0.0710) = -0.0218 +/- 0.0023
4331-30 (431-0)+(3-3) x2   (-0.0507) - (-0.0722) = +0.0215 +/- 0.0026    (+0.0478) - (+0.0710) = -0.0232 +/- 0.0025
4431-40 (431-0)+(4-4) x2   (-0.0772) - (-0.0722) = -0.0050 +/- 0.0023    (+0.0754) - (+0.0710) = +0.0044 +/- 0.0024
5431-50 (431-0)+(5-5) x1   (-0.0721) - (-0.0626) = -0.0095 +/- 0.0025    (+0.0717) - (+0.0615) = +0.0102 +/- 0.0025 

110-300 (11-30)+(0-0) x2   (+0.3093) - (+0.3603) = -0.0510 +/- 0.0028    (-0.3065) - (-0.3570) = +0.0505 +/- 0.0033
111-310 (11-30)+(1-1) x3   (+0.3769) - (+0.3821) = -0.0052 +/- 0.0037    (-0.3711) - (-0.3787) = +0.0076 +/- 0.0027
311-330 (11-30)+(3-3) x2   (+0.3072) - (+0.3602) = -0.0531 +/- 0.0035    (-0.3022) - (-0.3571) = +0.0550 +/- 0.0026
411-430 (11-30)+(4-4) x1   (+0.2259) - (+0.3120) = -0.0860 +/- 0.0030    (-0.2231) - (-0.3093) = +0.0862 +/- 0.0029  
511-530 (11-30)+(5-5) x1   (+0.2769) - (+0.3119) = -0.0351 +/- 0.0039    (-0.2710) - (-0.3093) = +0.0383 +/- 0.0027  

SUM OF SQUARES E42 (15):                  0.0179112                                         0.0187203

-- REDUCIBLE E33 --        

2211-00 (21-0)+(21-0) x1   (-0.1101) - (+0.0009) = -0.1109 +/- 0.0042    (+0.1165) - (+0.0009) = +0.1155 +/- 0.0054
3221-10 (21-0)+(32-1) below
4221-20 (21-0)+(42-2) below
5421-30 (21-0)+(54-3) x1   (-0.0459) - (+0.0023) = -0.0483 +/- 0.0040    (+0.0552) - (+0.0029) = +0.0523 +/- 0.0047
6521-40 (21-0)+(65-4) x1   (-0.0446) - (+0.0023) = -0.0469 +/- 0.0041    (+0.0532) - (+0.0029) = +0.0503 +/- 0.0043
7621-50 (21-0)+(76-5) x1   (-0.0285) - (+0.0022) = -0.0307 +/- 0.0040    (+0.0375) - (+0.0030) = +0.0345 +/- 0.0047
8721-60 (21-0)+(87-6) x1   (-0.0181) - (+0.0022) = -0.0203 +/- 0.0041    (+0.0251) - (+0.0031) = +0.0221 +/- 0.0039
9821-70 (21-0)+(98-7) x1   (-0.0105) - (+0.0022) = -0.0127 +/- 0.0038    (+0.0176) - (+0.0032) = +0.0144 +/- 0.0039

210-210 (21-0)+(0-21) ignore
211-320 (21-0)+(1-32) below
221-430 (21-0)+(2-43) x2   (+0.1863) - (+0.0034) = +0.1829 +/- 0.0050    (-0.1773) - (+0.0040) = -0.1813 +/- 0.0045
321-540 (21-0)+(3-54) x1   (+0.1150) - (+0.0023) = +0.1127 +/- 0.0036    (-0.1104) - (+0.0029) = -0.1132 +/- 0.0042
421-650 (21-0)+(4-65) x1   (+0.0482) - (+0.0023) = +0.0460 +/- 0.0038    (-0.0454) - (+0.0029) = -0.0483 +/- 0.0045
521-760 (21-0)+(5-76) x1   (+0.0306) - (+0.0022) = +0.0284 +/- 0.0039    (-0.0263) - (+0.0030) = -0.0294 +/- 0.0040
621-870 (21-0)+(6-87) x1   (+0.0186) - (+0.0022) = +0.0164 +/- 0.0037    (-0.0158) - (+0.0031) = -0.0189 +/- 0.0037
721-980 (21-0)+(7-98) x1   (+0.0119) - (+0.0022) = +0.0098 +/- 0.0040    (-0.0080) - (+0.0032) = -0.0112 +/- 0.0044

SUM OF SQUARES E33 (12):                   0.0677884                                       0.0698555  

-- REDUCIBLE  E43 --  

2111-300 (11-30)+(21-0) x6   (-0.0220) - (-0.0063) = -0.0157 +/- 0.0034    (+0.0110) - (-0.0065) = +0.0175 +/- 0.0032
3211-310 (11-30)+(32-1) below
4311-320 (11-30)+(43-2) below
5411-330 (11-30)+(54-3) x2   (+0.0070) - (-0.0034) = +0.0105 +/- 0.0031    (-0.0133) - (-0.0040) = -0.0092 +/- 0.0037
6511-430 (11-30)+(65-4) x1   (-0.0328) - (-0.0024) = -0.0305 +/- 0.0035    (+0.0259) - (-0.0029) = +0.0289 +/- 0.0033
7611-530 (11-30)+(76-5) x1   (-0.0160) - (-0.0023) = -0.0137 +/- 0.0034    (+0.0113) - (-0.0030) = +0.0143 +/- 0.0034
8711-630 (11-30)+(87-6) x1   (-0.0053) - (-0.0023) = -0.0030 +/- 0.0039    (-0.0009) - (-0.0031) = +0.0022 +/- 0.0035
9811-730 (11-30)+(98-7) x1   (-0.0057) - (-0.0023) = -0.0035 +/- 0.0036    (+0.0000) - (-0.0032) = +0.0032 +/- 0.0034

110-3210 (11-30)+(0-21) below
111-3320 (11-30)+(1-32) x6   (+0.0096) - (-0.0062) = +0.0157 +/- 0.0036    (-0.0200) - (-0.0067) = -0.0133 +/- 0.0031
211-4330 (11-30)+(2-43) below
311-5430 (11-30)+(3-54) below
411-6530 (11-30)+(4-65) x1   (-0.0388) - (-0.0024) = -0.0364 +/- 0.0035    (+0.0329) - (-0.0029) = +0.0358 +/- 0.0034
511-7630 (11-30)+(5-76) x1   (-0.0195) - (-0.0023) = -0.0172 +/- 0.0034    (+0.0150) - (-0.0030) = +0.0180 +/- 0.0035
611-8730 (11-30)+(6-87) x1   (-0.0056) - (-0.0023) = -0.0033 +/- 0.0034    (-0.0008) - (-0.0031) = +0.0023 +/- 0.0034
711-9830 (11-30)+(7-89) x1   (-0.0046) - (-0.0023) = -0.0023 +/- 0.0033    (+0.0001) - (-0.0032) = +0.0033 +/- 0.0035

2111-300 (11-30)+(21-0) below: (11-30)+(0-21) = 110-3210
2221-410 (22-41)+(21-0) below
3321-520 (33-52)+(21-0) below
4421-630 (44-63)+(21-0) x1   (+0.0226) - (-0.0024) = +0.0250 +/- 0.0036    (-0.0282) - (-0.0028) = -0.0253 +/- 0.0032
5521-740 (55-74)+(21-0) x1   (+0.0132) - (-0.0024) = +0.0156 +/- 0.0036    (-0.0189) - (-0.0029) = -0.0160 +/- 0.0032
6621-850 (66-85)+(21-0) x1   (+0.0097) - (-0.0023) = +0.0121 +/- 0.0035    (-0.0159) - (-0.0030) = -0.0129 +/- 0.0035
7721-960 (77-96)+(21-0) x1   (+0.0051) - (-0.0023) = +0.0074 +/- 0.0033    (-0.0109) - (-0.0031) = -0.0078 +/- 0.0036

3210-110 (30-11)+(21-0) below: 110-3210
4211-220 (41-22)+(21-0) below
5221-330 (52-33)+(21-0) x2   (-0.0626) - (-0.0035) = -0.0591 +/- 0.0039    (+0.0547) - (-0.0039) = +0.0587 +/- 0.0035
6321-440 (63-44)+(21-0) x1   (-0.0411) - (-0.0024) = -0.0386 +/- 0.0040    (+0.0338) - (-0.0028) = +0.0366 +/- 0.0033
7421-550 (74-55)+(21-0) x1   (-0.0215) - (-0.0024) = -0.0192 +/- 0.0034    (+0.0158) - (-0.0029) = +0.0188 +/- 0.0038
8521-660 (85-66)+(21-0) x1   (-0.0120) - (-0.0023) = -0.0097 +/- 0.0034    (+0.0065) - (-0.0030) = +0.0095 +/- 0.0035
9621-770 (96-77)+(21-0) x1   (-0.0085) - (-0.0023) = -0.0062 +/- 0.0033    (+0.0036) - (-0.0031) = +0.0067 +/- 0.0037

32221-00 (322-0)+(21-0) x6   (-0.0360) - (+0.0023) = -0.0383 +/- 0.0039    (+0.0396) - (+0.0026) = +0.0370 +/- 0.0031
33222-10 (322-0)+(32-1) x6   (-0.0502) - (+0.0023) = -0.0524 +/- 0.0029    (+0.0561) - (+0.0026) = +0.0535 +/- 0.0040
43322-20 (322-0)+(43-2) below
54322-30 (322-0)+(54-3) below 
65322-40 (322-0)+(65-4) x1   (+0.0059) - (+0.0009) = +0.0050 +/- 0.0034    (-0.0036) - (+0.0012) = -0.0048 +/- 0.0031
76322-50 (322-0)+(76-5) x1   (+0.0083) - (+0.0009) = +0.0075 +/- 0.0032    (-0.0065) - (+0.0012) = -0.0077 +/- 0.0033
87322-60 (322-0)+(87-6) x1   (+0.0013) - (+0.0008) = +0.0004 +/- 0.0037    (+0.0005) - (+0.0012) = -0.0007 +/- 0.0034
98322-70 (322-0)+(98-7) x1   (+0.0017) - (+0.0008) = +0.0009 +/- 0.0034    (+0.0004) - (+0.0013) = -0.0009 +/- 0.0039

3220-210 (322-0)+(0-21) below
3221-320 (322-0)+(1-32) below
3222-430 (322-0)+(2-43) below
3322-540 (322-0)+(3-54) x2   (+0.0156) - (+0.0013) = +0.0143 +/- 0.0041    (-0.0157) - (+0.0016) = -0.0173 +/- 0.0037
4322-650 (322-0)+(4-65) x1   (-0.0122) - (+0.0009) = -0.0130 +/- 0.0040    (+0.0137) - (+0.0012) = +0.0126 +/- 0.0032
5322-760 (322-0)+(5-76) x1   (+0.0037) - (+0.0009) = +0.0029 +/- 0.0038    (-0.0035) - (+0.0012) = -0.0047 +/- 0.0037
6322-870 (322-0)+(6-87) x1   (+0.0004) - (+0.0008) = -0.0005 +/- 0.0035    (+0.0012) - (+0.0012) = +0.0000 +/- 0.0034
7322-980 (322-0)+(7-98) x1   (+0.0018) - (+0.0008) = +0.0009 +/- 0.0035    (-0.0002) - (+0.0013) = -0.0014 +/- 0.0037

32221-00 (322-0)+(21-0) below: (322-0)+(0-21) = 3220-210
43321-10 (433-1)+(21-0) below
54421-20 (544-2)+(21-0) below
65521-30 (655-3)+(21-0) x1   (-0.0141) - (+0.0009) = -0.0150 +/- 0.0037    (+0.0183) - (+0.0011) = +0.0172 +/- 0.0038
76621-40 (766-4)+(21-0) x1   (-0.0141) - (+0.0009) = -0.0149 +/- 0.0035    (+0.0180) - (+0.0012) = +0.0168 +/- 0.0034
87721-50 (877-5)+(21-0) x1   (-0.0097) - (+0.0009) = -0.0105 +/- 0.0037    (+0.0133) - (+0.0012) = +0.0121 +/- 0.0038
98821-60 (988-6)+(21-0) x1   (-0.0059) - (+0.0009) = -0.0068 +/- 0.0036    (+0.0095) - (+0.0012) = +0.0083 +/- 0.0030

210-3220 (0-322)+(21-0) below: 3220-210
211-4330 (1-433)+(21-0) below
221-5440 (2-544)+(21-0) x2   (+0.0533) - (+0.0013) = +0.0520 +/- 0.0034    (-0.0544) - (+0.0016) = -0.0560 +/- 0.0037
321-6550 (3-655)+(21-0) x1   (+0.0360) - (+0.0009) = +0.0351 +/- 0.0035    (-0.0349) - (+0.0011) = -0.0360 +/- 0.0035
421-7660 (4-766)+(21-0) x1   (+0.0153) - (+0.0009) = +0.0144 +/- 0.0036    (-0.0150) - (+0.0012) = -0.0162 +/- 0.0039
521-8770 (5-877)+(21-0) x1   (+0.0115) - (+0.0009) = +0.0107 +/- 0.0035    (-0.0105) - (+0.0012) = -0.0116 +/- 0.0037
621-9880 (6-988)+(21-0) x1   (+0.0073) - (+0.0009) = +0.0065 +/- 0.0035    (-0.0066) - (+0.0012) = -0.0079 +/- 0.0034

43211-00 (431-0)+(21-0) x4   (-0.0148) - (+0.0010) = -0.0159 +/- 0.0035    (+0.0169) - (+0.0011) = +0.0159 +/- 0.0034
43321-10 (431-0)+(32-1) below
44331-20 (431-0)+(43-2) x4   (-0.0044) - (+0.0010) = -0.0054 +/- 0.0034    (+0.0073) - (+0.0011) = +0.0062 +/- 0.0035
54431-30 (431-0)+(54-3) below
65431-40 (431-0)+(65-4) below
76431-50 (431-0)+(76-5) x1   (-0.0080) - (+0.0005) = -0.0084 +/- 0.0036    (+0.0083) - (+0.0006) = +0.0077 +/- 0.0038
87431-60 (431-0)+(87-6) x1   (+0.0035) - (+0.0005) = +0.0030 +/- 0.0033    (-0.0023) - (+0.0006) = -0.0029 +/- 0.0042
98431-70 (431-0)+(98-7) x1   (+0.0007) - (+0.0005) = +0.0002 +/- 0.0033    (-0.0002) - (+0.0006) = -0.0008 +/- 0.0033

4310-210 (431-0)+(0-21) below
4311-320 (431-0)+(1-32) below
4321-430 (431-0)+(2-43) below
4331-540 (431-0)+(3-54) below
4431-650 (431-0)+(4-65) x2   (+0.0056) - (+0.0007) = +0.0049 +/- 0.0040    (-0.0046) - (+0.0008) = -0.0054 +/- 0.0031
5431-760 (431-0)+(5-76) x1   (-0.0120) - (+0.0005) = -0.0125 +/- 0.0038    (+0.0137) - (+0.0006) = +0.0132 +/- 0.0034
6431-870 (431-0)+(6-87) x1   (+0.0021) - (+0.0005) = +0.0016 +/- 0.0036    (-0.0011) - (+0.0006) = -0.0017 +/- 0.0038
7431-980 (431-0)+(7-98) x1   (+0.0004) - (+0.0005) = -0.0001 +/- 0.0041    (+0.0001) - (+0.0006) = -0.0005 +/- 0.0032

43211-00 (431-0)+(21-0) below: (431-0)+(0-21) = 4310-210
54221-10 (542-1)+(21-0) below
65321-20 (653-2)+(21-0) below
76421-30 (764-3)+(21-0) x1   (-0.0206) - (+0.0005) = -0.0211 +/- 0.0039    (+0.0229) - (+0.0006) = +0.0223 +/- 0.0033
87521-40 (875-4)+(21-0) x1   (-0.0168) - (+0.0005) = -0.0172 +/- 0.0035    (+0.0188) - (+0.0006) = +0.0183 +/- 0.0043
98621-50 (986-5)+(21-0) x1   (-0.0113) - (+0.0005) = -0.0117 +/- 0.0034    (+0.0136) - (+0.0006) = +0.0131 +/- 0.0034

210-4310 (0-431)+(21-0) below: (431-0)+(0-21) = 4310-210               
211-5420 (1-542)+(21-0) below
221-6530 (2-653)+(21-0) x1   (+0.0648) - (+0.0004) = +0.0645 +/- 0.0040    (-0.0613) - (+0.0004) = -0.0617 +/- 0.0037
321-7640 (3-764)+(21-0) x1   (+0.0381) - (+0.0005) = +0.0376 +/- 0.0034    (-0.0368) - (+0.0006) = -0.0374 +/- 0.0036
421-8750 (4-875)+(21-0) x1   (+0.0203) - (+0.0005) = +0.0198 +/- 0.0037    (-0.0198) - (+0.0006) = -0.0204 +/- 0.0034
521-9860 (5-968)+(21-0) x1   (+0.0122) - (+0.0005) = +0.0118 +/- 0.0037    (-0.0117) - (+0.0006) = -0.0123 +/- 0.0034

SUM OF SQUARES E43 (56):                       0.0270419                                     0.0274663 

-- REDUCIBLE  E52 -- 

43320-00 (4332-0)+(0-0) x2   (-0.0453) - (-0.0586) = +0.0132 +/- 0.0023    (+0.0452) - (+0.0590) = -0.0138 +/- 0.0019
43321-10 (4332-0)+(1-1)   below
43322-20 (4332-0)+(2-2)   below
43332-30 (4332-0)+(3-3) x3   (-0.0466) - (-0.0621) = +0.0156 +/- 0.0017    (+0.0474) - (+0.0626) = -0.0152 +/- 0.0019
44332-40 (4332-0)+(4-4) x2   (-0.0481) - (-0.0586) = +0.0104 +/- 0.0022    (+0.0484) - (+0.0590) = -0.0107 +/- 0.0018
54332-50 (4332-0)+(5-5) x1   (-0.0398) - (-0.0507) = +0.0109 +/- 0.0023    (+0.0395) - (+0.0511) = -0.0116 +/- 0.0022
64332-60 (4332-0)+(6-6) x1   (-0.0527) - (-0.0507) = -0.0020 +/- 0.0023    (+0.0529) - (+0.0511) = +0.0018 +/- 0.0019

54220-00 (5422-0)+(0-0) x2   (-0.0338) - (-0.0443) = +0.0104 +/- 0.0021    (+0.0371) - (+0.0481) = -0.0110 +/- 0.0019
54221-10 (5422-0)+(1-1)   below
54222-20 (5422-0)+(2-2) x3   (-0.0460) - (-0.0469) = +0.0009 +/- 0.0018    (+0.0500) - (+0.0510) = -0.0010 +/- 0.0017
54322-30 (5422-0)+(3-3)   below                     
54422-40 (5422-0)+(4-4) x2   (-0.0368) - (-0.0443) = +0.0074 +/- 0.0021    (+0.0408) - (+0.0481) = -0.0074 +/- 0.0021
55422-50 (5422-0)+(5-5) x2   (-0.0359) - (-0.0443) = +0.0083 +/- 0.0021    (+0.0385) - (+0.0481) = -0.0097 +/- 0.0021
65422-60 (5422-0)+(6-6) x2   (-0.0240) - (-0.0766) = +0.0526 +/- 0.0036    (+0.0265) - (+0.0834) = -0.0569 +/- 0.0045

54410-00 (5441-0)+(0-0) x2   (-0.0096) - (-0.0140) = +0.0044 +/- 0.0020    (+0.0080) - (+0.0124) = -0.0044 +/- 0.0021
54411-10 (5441-0)+(1-1) x2   (-0.0054) - (-0.0140) = +0.0087 +/- 0.0023    (+0.0046) - (+0.0124) = -0.0078 +/- 0.0020
54421-20 (5441-0)+(2-2)   below
54431-30 (5441-0)+(3-3)   below
54441-40 (5441-0)+(4-4) x3   (-0.0170) - (-0.0149) = -0.0022 +/- 0.0017    (+0.0154) - (+0.0132) = +0.0022 +/- 0.0018
55441-50 (5441-0)+(5-5) x2   (-0.0176) - (-0.0140) = -0.0036 +/- 0.0021    (+0.0155) - (+0.0124) = +0.0031 +/- 0.0020
65441-60 (5441-0)+(6-6) x1   (-0.0096) - (-0.0122) = +0.0026 +/- 0.0024    (+0.0068) - (+0.0108) = -0.0039 +/- 0.0030

65310-00 (6531-0)+(0-0) x2   (-0.0130) - (-0.0176) = +0.0046 +/- 0.0022    (+0.0142) - (+0.0183) = -0.0042 +/- 0.0022
65311-10 (6531-0)+(1-1) x2   (-0.0084) - (-0.0176) = +0.0092 +/- 0.0023    (+0.0086) - (+0.0183) = -0.0098 +/- 0.0021
65321-20 (6531-0)+(2-2)   below
65331-30 (6531-0)+(3-3) x2   (-0.0174) - (-0.0176) = +0.0002 +/- 0.0019    (+0.0175) - (+0.0184) = -0.0008 +/- 0.0019
65431-40 (6531-0)+(4-4)   below
65531-50 (6531-0)+(5-5) x2   (-0.0197) - (-0.0176) = -0.0021 +/- 0.0020    (+0.0208) - (+0.0184) = +0.0024 +/- 0.0022
66531-60 (6531-0)+(6-6) x2   (-0.0134) - (-0.0176) = +0.0042 +/- 0.0020    (+0.0135) - (+0.0184) = -0.0048 +/- 0.0020

4110-200 (411-20)+(0-0) x2   (-0.1075) - (-0.1107) = +0.0032 +/- 0.0023    (+0.1024) - (+0.1045) = -0.0021 +/- 0.0024
4111-210 (411-20)+(1-1) x3   (-0.1011) - (-0.1174) = +0.0164 +/- 0.0019    (+0.0936) - (+0.1109) = -0.0173 +/- 0.0019
4211-220 (411-20)+(2-2)   below
4311-320 (411-20)+(3-3)   below
4411-420 (411-20)+(4-4) x2   (-0.0911) - (-0.1107) = +0.0196 +/- 0.0020    (+0.0867) - (+0.1046) = -0.0179 +/- 0.0024
5411-520 (411-20)+(5-5) x1   (-0.0729) - (-0.0959) = +0.0230 +/- 0.0026    (+0.0684) - (+0.0906) = -0.0222 +/- 0.0028
6411-620 (411-20)+(6-6) x1   (-0.0919) - (-0.0959) = +0.0040 +/- 0.0028    (+0.0862) - (+0.0906) = -0.0044 +/- 0.0023

3310-500 (331-50)+(0-0) x2   (-0.0170) - (-0.0120) = -0.0050 +/- 0.0022    (+0.0174) - (+0.0118) = +0.0057 +/- 0.0020
3311-510 (331-50)+(1-1) x2   (-0.0246) - (-0.0120) = -0.0126 +/- 0.0024    (+0.0219) - (+0.0118) = +0.0101 +/- 0.0021
3321-520 (331-50)+(2-2)   below
3331-530 (331-50)+(3-3) x3   (-0.0187) - (-0.0127) = -0.0059 +/- 0.0017    (+0.0184) - (+0.0125) = +0.0059 +/- 0.0018
4331-540 (331-50)+(4-4)   below
5331-550 (331-50)+(5-5) x2   (-0.0079) - (-0.0120) = +0.0041 +/- 0.0020    (+0.0076) - (+0.0118) = -0.0042 +/- 0.0020
6331-650 (331-50)+(6-6) x1   (+0.0005) - (-0.0104) = +0.0109 +/- 0.0027    (-0.0007) - (+0.0102) = -0.0109 +/- 0.0023

2220-400 (222-40)+(0-0) x2   (-0.0465) - (-0.0289) = -0.0176 +/- 0.0025    (+0.0443) - (+0.0289) = +0.0154 +/- 0.0018
2221-410 (222-40)+(1-1)   below
2222-420 (222-40)+(2-2) x4   (-0.0406) - (-0.0316) = -0.0090 +/- 0.0016    (+0.0406) - (+0.0317) = +0.0089 +/- 0.0018
3222-430 (222-40)+(3-3)   below
4222-440 (222-40)+(4-4) x2   (-0.0245) - (-0.0289) = +0.0044 +/- 0.0024    (+0.0240) - (+0.0289) = -0.0050 +/- 0.0020
5222-540 (222-40)+(5-5) x1   (-0.0013) - (-0.0250) = +0.0238 +/- 0.0020    (+0.0014) - (+0.0251) = -0.0236 +/- 0.0023
6222-640 (222-40)+(6-6) x1   (-0.0166) - (-0.0250) = +0.0084 +/- 0.0026    (+0.0168) - (+0.0251) = -0.0083 +/- 0.0025

110-5400 (11-540)+(0-0) x2   (+0.0996) - (+0.1202) = -0.0206 +/- 0.0020    (-0.0999) - (-0.1192) = +0.0193 +/- 0.0023
111-5410 (11-540)+(1-1) x3   (+0.1219) - (+0.1274) = -0.0055 +/- 0.0020    (-0.1201) - (-0.1264) = +0.0063 +/- 0.0021
211-5420 (11-540)+(2-2)   below
311-5430 (11-540)+(3-3)   below 
411-5440 (11-540)+(4-4) x2   (+0.0949) - (+0.1201) = -0.0252 +/- 0.0027    (-0.0927) - (-0.1192) = +0.0265 +/- 0.0019
511-5540 (11-540)+(5-5) x2   (+0.1042) - (+0.1201) = -0.0159 +/- 0.0022    (-0.1024) - (-0.1193) = +0.0169 +/- 0.0020
611-6540 (11-540)+(6-6) x1   (+0.0842) - (+0.1040) = -0.0199 +/- 0.0025    (-0.0832) - (-0.1033) = +0.0201 +/- 0.0027
                     
SUM OF SQUARES E52 (40):                    0.00824359                              0.00864855

--------------------------------------------------------------------------------------------------------------------
MOMENTS WITH MULTIPLE DECOMPOSITION
--------------------------------------------------------------------------------------------------------------------

                                  DIRECT                               INVERSE

-- REDUCIBLE  E42.33 --

3221-10            (-0.2099) - (-0.1140) = -0.0959         (+0.2207) - (+0.1235) = +0.0972 
  (21-0)+(32-1) x2           - (+0.0024)                             - (+0.0027)
  (322-0)+(1-1) x1           - (-0.1164)                             - (+0.1208)

4321-20            (-0.0758) - (-0.0609) = -0.0149         (+0.0775) - (+0.0635) = +0.0140 
  (21-0)+(43-2) x1           - (+0.0017)                             - (+0.0020)
  (431-0)+(2-2) x1           - (-0.0626)                             - (+0.0615)

211-320            (+0.3494) - (+0.3144) = +0.0350         (-0.3367) - (-0.3065) = -0.0302
  (21-0)+(1-32) x2           - (+0.0024)                             - (+0.0027)
  (11-30)+(2-2) x1           - (+0.3120)                             - (-0.3092)

SUM OF SQUARES E42.33 (3):                0.0106438                               0.0105559  


-- REDUCIBLE  E43.52 --   

211-5420           (+0.1006) - (+0.1045) = -0.0039        (-0.0993) - (-0.1027) = 0.0034
  (1-542)+(21-0) x2          - (+0.0005)                            - (+0.0005)
  (11-540)+(2-2) x1          - (+0.1040)                            - (-0.1032)
            
2221-410            (+0.0042) - (-0.0281) = +0.0323       (-0.0080) - (+0.0218) = -0.0298
  (22-41)+(21-0) x3           - (-0.0031)                           - (-0.0033)
  (222-40)+(1-1) x1           - (-0.0250)                           - (+0.0251)

311-5430            (+0.0896) - (+0.1023) = -0.0127       (-0.0897) - (-0.1053) = +0.0156
  (11-30)+(3-54) x1            - (-0.0017)                          - (-0.0020)
  (11-540)+(3-3) x1            - (+0.1040)                          - (-0.1033)

3222-430           (+0.0087) - (-0.0239) = +0.0326        (-0.0087) - (+0.0264) = -0.0351
  (322-0)+(2-43) x3          - (+0.0011)                            - (+0.0013)
  (222-40)+(3-3) x1          - (-0.0250)                            - (+0.0251)

3321-520            (-0.0056) - (-0.0121) = +0.0065       (+0.0025) - (+0.0082) = -0.0057
  (33-52)+(21-0) x1           - (-0.0017)                           - (-0.0020)
  (331-50)+(2-2) x1           - (-0.0104)                           - (+0.0102)

4211-220            (-0.1106) - (-0.1128) = +0.0022       (+0.1010) -  (+0.1024) = -0.0014
  (41-22)+(21-0) x2           - (-0.0021)                           - (-0.0022)
  (411-20)+(2-2) x2           - (-0.1107)                           - (+0.1046)

4331-540           (+0.0178) - (-0.0099) = +0.0277        (-0.0170) - (+0.0108) = -0.0278
  (431-0)+(3-54) x2          - (+0.0005)                            - (+0.0006)
  (331-50)+(4-4) x1          - (-0.0104)                            - (+0.0102)

43322-20            (-0.0491) - (-0.0579) = +0.0088       (+0.0481) - (+0.0599) = -0.0118
  (322-0)+(43-2) x2           - (+0.0007)                           - (+0.0009)
  (4332-0)+(2-2) x2           - (-0.0586)                           - (+0.0590)

54221-10           (-0.0651) - (-0.0378) = -0.0273        (+0.0693) - (+0.0422) = 0.0271
  (542-1)+(21-0) x2          - (+0.0005)                            - (+0.0005)
  (5422-0)+(1-1) x1          - (-0.0383)                            - (+0.0417)

54322-30           (-0.0315) - (-0.0377) = +0.0062        (+0.0343) - (+0.0425) = -0.0082
  (322-0)+(54-3) x1          - (+0.0006)                            - (+0.0008)
  (5422-0)+(3-3) x1          - (-0.0383)                            - (+0.0417)

54421-20           (-0.0095) - (-0.0116) = +0.0021        (+0.0095) - (+0.0115) = -0.0020   
  (544-2)+(21-0) x1          - (+0.0006)                            - (+0.0008)
  (5441-0)+(2-2) x1          - (-0.0122)                            - (+0.0107)

54431-30           (-0.0083) - (-0.0117) = +0.0034        (+0.0065) - (+0.0113) = -0.0048
  (431-0)+(54-3) x2          - (+0.0005)                            - (+0.0006)
  (5441-0)+(3-3) x1          - (-0.0122)                            - (+0.0107)

65321-20           (-0.0226) - (-0.0148) = -0.0078        (+0.0244) - (+0.0163) = 0.0081
  (653-2)+(21-0) x1          - (+0.0004)                            - (+0.0004)
  (6531-0)+(2-2) x1          - (-0.0152)                            - (+0.0159)

65431-40           (-0.0165) - (-0.0149) = -0.0016        (+0.0175) - (+0.0163) = +0.0012
  (431-0)+(65-4) x1          - (+0.0003)                            - (+0.0004)
  (6531-0)+(4-4) x1          - (-0.0152)                            - (+0.0159)


SUM OF SQUARES E43.52 (14):              0.00403747                             0.00421724 

-- REDUCIBLE  E43.43 --   

211-4330           (+0.0022) - (-0.0022) = +0.0044        (-0.0056) - (-0.0024) = -0.0032
  (11-30)+(2-43) x2          - (-0.0035)                            - (-0.0039)
  (1-433)+(21-0) x2          - (+0.0013)                            - (+0.0015)

SUM OF SQUARES E43.43 (1):               0.00001936                             0.00001024

-- REDUCIBLE  E43.43.52 --   

4311-320           (-0.0633) - (-0.0971) = +0.0338        (+0.0554) - (+0.0891) = -0.0337
  (431-0)+(1-32) x2          - (+0.0005)                            - (+0.0005)
  (11-30)+(43-2) x1          - (-0.0017)                            - (-0.0020)
  (411-20)+(3-3) x1          - (-0.0959)                            - (+0.0906)

43321-10           (-0.0792) - (-0.0495) = -0.0297        (+0.0831) - (+0.0525) = +0.0306
  (433-1)+(21-0) x1          - (+0.0007)                            - (+0.0008)
  (431-0)+(32-1) x2          - (+0.0005)                            - (+0.0005)
  (4332-0)+(1-1) x1          - (-0.0507)                            - (+0.0511)

SUM OF SQUARES E43.43.52 (2):             0.00202453                            0.00207205   

-- REDUCIBLE  E43.322 --   

3211-310           (-0.2078) - (-0.2493) = +0.0415        (+0.2073) - (+0.2512) = -0.0439 
  (11-30)+(32-1)     x1      - (-0.0010)                            - (-0.0011)
  (3-3)+(1-1)+(21-0) x2      - (-0.2483)                            - (+0.2523)

110-3210           (-0.2710) - (-0.2460) = -0.0250        (+0.2801) - (+0.2548) = +0.0253
  (11-30)+(0-21)     x1      - (-0.0010)                            - (-0.0011)
  (1-1)+(0-0)+(1-32) x2      - (-0.2450)                            - (+0.2559)

3220-210           (-0.2811) - (-0.2446) = -0.0365        (+0.2920) - (+0.2563) = +0.0357
  (322-0)+(0-21)     x1      - (+0.0004)                            - (+0.0004)
  (2-2)+(0-0)+(32-1) x2      - (-0.2450)                            - (+0.2559)
            
3221-320           (-0.1656) - (-0.2479) = +0.0823        (+0.1654) - (+0.2527) = -0.0873
  (322-0)+(1-32)     x1      - (+0.0004)                            - (+0.0004)
  (3-3)+(2-2)+(21-0) x2      - (-0.2483)                            - (+0.2523)
            
4310-210           (-0.1790) - (-0.2092) = +0.0302        (+0.1989) - (+0.2253) = -0.0264
  (431-0)+(0-21)     x1      - (+0.0003)                            - (+0.0003)
  (1-1)+(0-0)+(43-2) x1      - (-0.2095)                            - (+0.2250)
            
4321-430           (-0.1591) - (-0.2148) = +0.0557        (+0.1587) - (+0.2188) = -0.0601
  (431-0)+(2-43)     x1      - (+0.0002)                            - (+0.0003)
  (4-4)+(3-3)+(21-0) x1      - (-0.2150)                            - (+0.2185)

SUM OF SQUARES E43.322 (6):               0.0144673                             0.015772



--------------------------------------------------------
Moments of order 9 and 12 built from flux J
(direct cascade only)
--------------------------------------------------------

m=9, "I" stands for conj(J), last character is shift

 JJJ0:  +0.0034 +/-  0.0039      0 "bb^*" pairs
 IJJ0:  -0.1953 +/-  0.0055      3
 IIJ0:  -0.1953 +/-  0.0055      3
 JIJ0:  -0.1953 +/-  0.0055      3

 JJJ1:  -0.0511 +/-  0.0040      2
 IJJ1:  +0.0181 +/-  0.0039      2
 JIJ1:  -0.0491 +/-  0.0048      2
 IIJ1:  +0.0075 +/-  0.0038      1

 JJJ2:  -0.0166 +/-  0.0033      2
 IJJ2:  +0.0011 +/-  0.0039      1
 JIJ2:  +0.0016 +/-  0.0034      0
 IIJ2:  +0.0350 +/-  0.0042      1

 JJJ3:  +0.0067 +/-  0.0034      0
 IJJ3:  +0.0062 +/-  0.0037      0
 JIJ3:  -0.0090 +/-  0.0040      0
 IIJ3:  -0.0095 +/-  0.0039      0

m=12, no shift:
 JJJJ:  +0.0002 +/-  0.0037


--------------------------------------------------------
\end{verbatim}
\end{footnotesize}

In the last portion of the table above we consider 9th (and one 12th) order moments
made out of shifted triplets. Here, {\tt J} stand for $J =
b_{i-2}b_{i-1}b_i^*$, while {\tt I} stands for $J^*$, and the last
charecter indicates the shift between triplets. So, {\tt IIJ2} means
$J^*_{i-4}J^*_{i-2}J_i$.  The table shows simple computed averages
with no subtraction.  We can evaluate substracted components for
moments with decomposition $m=9=2+2+5$, such as {\tt JIJ1} or
$J_{i-2}J^*_{i-1}J_i$: $43\bar{2} + \bar{3}\bar{2}1 + 21\bar{0} =
2\bar{2} + 3\bar{3} + 4\bar{2}11\bar{0}$.  The set does not account for every shift with
overlap. In particalar, the combinations with non-equal shifts, such
as $J_{i-3}J_{i-1}J_i$, are not considered.